\documentclass[final,3p,times]{elsarticle}

\usepackage{subfigure,color}
\usepackage{hypernat}
\usepackage{multirow}
\usepackage{booktabs}
\usepackage{threeparttable}
\usepackage{makecell}
\usepackage{epstopdf}
\usepackage{float}
\usepackage{appendix}

\usepackage{amsmath,amssymb}
\usepackage{lipsum}
\usepackage{bm}
\usepackage{graphicx}
\usepackage{graphics}
\usepackage{amsmath}
\usepackage{array}
\usepackage{hyperref}
\usepackage[capitalise]{cleveref}
\usepackage{textcomp}
\usepackage{epstopdf}
\usepackage{cases}
\usepackage{amssymb}
\usepackage{multirow}
\usepackage{siunitx}
\usepackage{cases}

\numberwithin{equation}{section}

\graphicspath{{Figs/}}

\begin{document}

\title{Modeling Heat Conduction with Dual-Dissipative Variables: A Mechanism-Data Fusion Method}
\author[1]{Leheng Chen}
\ead{chenlh@pku.edu.cn}
\author[add1]{Chuang Zhang\corref{cor1}}
\ead{zhangc520@hdu.edu.cn}
\author[6,7]{Jin Zhao\corref{cor1}}
\ead{zjin@cnu.edu.cn}
\address[1]{School of Mathematical Sciences, Peking University, Beijing, 100871, China}
\address[add1]{Department of Physics, Hangzhou Dianzi University, Hangzhou 310018, China}
\address[6]{Academy for Multidisciplinary Studies, Capital Normal University, Beijing, 100048, China}
\address[7]{Beijing National Center for Applied Mathematics, Beijing, 100048, China}
\cortext[cor1]{Corresponding author}
%

\date{\today}

\begin{abstract}

Many macroscopic non-Fourier heat conduction models have been developed in the past decades based on Chapman-Enskog, Hermite or other small perturbation expansion methods. These macroscopic models have made great success on capturing non-Fourier thermal behaviors in solid materials, but most of them are limited by small Knudsen numbers and incapable of capturing highly non-equilibrium or ballistic thermal transport. In this paper, we provide a new strategy for constructing macroscopic non-Fourier heat conduction modeling, that is, using data-driven deep learning methods combined with non-equilibrium thermodynamics instead of small perturbation expansion. We present the mechanism-data fusion method, an approach that seamlessly integrates the rigorous framework of Conservation-Dissipation Formalism (CDF) with the flexibility of machine learning to model non-Fourier heat conduction. Leveraging the conservation-dissipation principle with dual-dissipative variables, we derive an interpretable series of partial differential equations, fine-tuned through a training strategy informed by data from the phonon Boltzmann transport equation. Moreover, we also present the inner-step operation to narrow the gap from the discrete form to the continuous system. Through numerical tests, our model demonstrates excellent predictive capabilities across various heat conduction regimes, including diffusive, hydrodynamic and ballistic regimes, and displays its robustness and precision even with discontinuous initial conditions.

\end{abstract}

\maketitle

\section{Introduction}

{\color{black}{Heat conduction in solid materials is generally described by the Fourier's law in our daily life, which implies a diffusive phonon transport process~\cite{2005Nanoscale,chen_non-fourier_2021}.
The heat flux is proportional to the temperature gradient and the bulk thermal conductivity only depends on materials' components and temperature in this classic empirical formula.
{\color{black}{However, with the rapid development of advanced manufacturing, ultrafast lasers, nanomaterials and other technologies~\cite{2005Nanoscale,cahill2003nanoscale,cahill2014nanoscale}, specifically, the laser response time has been shortened from microseconds to picoseconds or even femtoseconds~\cite{RevModPhysJoseph89}, and the emergence of low-dimensional materials~\cite{lee_hydrodynamic_2015,huberman_observation_2019,hydrodynamics_review_2022} such as graphene and carbon nanotubes, typical macroscopic models or theories are difficult to accurately describe the thermal transport phenomenon at the micro/nano scale~\cite{GUO20151,PRM076001}.  }}
When the system characteristic length/time is comparable to or smaller than phonon mean free path/relaxation time, or the momentum-conserved normal scattering process dominates heat conduction, the Fourier's law will be broken. 
Lots of non-Fourier heat conduction phenomena have been found \cite{GUO20151,PRM076001,cahill2003nanoscale,cahill2014nanoscale,2005Nanoscale,CZhang2022JAP}, for example wave like propagation of heat with finite speed or heat wave~\cite{CZhang2022JAP,RevModPhysJoseph89}, size effects~\cite{cahill2003nanoscale,cahill2014nanoscale} and hydrodynamic phonon transport with sufficient normal scattering process~\cite{lee_hydrodynamic_2015,huberman_observation_2019,hydrodynamics_review_2022}. 
To study the non-Fourier heat conduction processes in solid materials, many theoretical/numerical methods have been developed as well as experimental measuring techniques \cite{cahill2003nanoscale,cahill2014nanoscale,PhysRev.148.766,PhysRevX.10.011019,esee8c146,huberman_observation_2019,chen_non-fourier_2021}.
Compared with microscopic or mesoscopic methods, macroscopic equations have fewer degrees of freedom, higher computational efficiency, and can efficiently solving the engineering multi-scale heat transfer problems by introducing some empirical correction terms or coefficients so that they have more audience until today and many researchers are still keen on macroscopic non-Fourier heat conduction modeling~\cite{cattaneo1948sulla,gurtin_general_1968,RevModPhysJoseph89,PhysRevB.103.L140301}.}}

{\color{black}{In the past decades, many macroscopic heat conduction models have been developed for describing non-Fourier heat conduction phenomena by introducing phase lag, nonlocal, nonlinear, fractional or other complex high-order terms to reflect the actual relationship between heat flux and temperature at the micro/nano scales \cite{GUO20151,PhysRevB.103.L140301,li_beyond_2020,cattaneo1948sulla,PhysRevB.98.104304,xu_modeling_2012}.
However, the task of building models that capture the fundamental aspects of underlying physics in simple, understandable, and reliably universal forms continues to be a complex and challenging endeavor \cite{dawning2021}. 
This challenge is dual-faceted: On one hand, the deliberate pace of theoretical advancement in physics encumbers rapid modeling innovation; on the other hand, rigid physical axioms severely constrain the creation of computationally feasible models which, compounded by the inevitable distortions wrought by mathematical abstractions, circumscribe their applicability. For example, many macroscopic non-Fourier heat conduction models are derived by the low-order Chapman-Enskog expansion of the Boltzmann transport equation, which requires that the system characteristic size/time should be larger than mean free path/relaxation time \cite{GUO20151}. }}

{\color{black}{In this paper, we provide a new strategy for constructing macroscopic non-Fourier heat conduction modeling, that is, using data-driven deep learning methods \cite{e_machine-learning-assisted_2021,han2020integrating,han2019uniformly,HuangMaZhouYong2021,ZHAO2022122396} combined with non-equilibrium thermodynamics instead of typical small perturbation expansion\cite{cattaneo1948sulla,gurtin_general_1968,RevModPhysJoseph89,PhysRevB.103.L140301}.
Firstly, theoretically deriving a macroscopic heat conduction equation with unknown functions or parameters which is valid at any scale, and secondly using data-driven deep learning methods to train and learn these unknown functions or parameters.
We present the mechanism-data fusion method, an {\color{black}innovative} approach that seamlessly integrates the rigorous framework of Conservation-Dissipation Formalism (CDF) with the flexibility of machine learning to model non-Fourier heat conduction.  
Through comprehensive numerical tests, our model demonstrates superior predictive capabilities across various heat conduction regimes, including diffusive, hydrodynamic and ballistic regimes, and displays its robustness and precision even with discontinuous initial conditions.}}

The paper is organized as follows. The theoretical derivations of the heat conduction model with two dissipative variables are presented in Sec.~\ref{sec:model}. In Sec.~\ref{sec:learnAI}, we show how to learn the unknown functions and optimize parameters, and the ISO method is presented here. Additionally, the training data generated by the phonon Boltzmann transport equation (BTE) is introduced in this section too. Lots of numerical tests and discussions are conducted to validate our model in Sec.~\ref{sec:numericalresults} and~\ref{sec:analysis}. Conclusions and remarks are given in Sec.~\ref{sec:conclusion}. 

\section{Model}
\label{sec:model}

In this section we use the framework of Conservation-Dissipation Formalism (CDF) to derive the model of heat conduction with two dissipative variables, and the CDF can guarantee that this model naturally satisfies the first and second laws of thermodynamics. 
With reasonable assumption for the entropy of the system, we can obtain the specific form of the model.

\subsection{Model Derivation}
{\color{black}{For the heat conduction in solid materials without external heat source}}, the first law of thermodynamics reads as $\partial_t u + \nabla \cdot {\bf{q}} = 0$, where $u$ is the internal energy, and ${\bf{q}}$ is the corresponding heat flux. 
From the mathematical point of view, Fourier's law is ${\bf{q}}=-\kappa  \nabla \theta$ with bulk thermal conductivity $\kappa$ and temperature $\theta$.
The goal of this work is devoted to integrating the mechanism and data to model the non-Fourier's heat conduction. To begin with, we present the derivation of the model by using CDF. Specifically, we introduce two dissipative variables $\bf{w},\bf{Q}$ whose counterpart is the conservative variable, $u$; $\bf{w}$ is a vector which has the same dimension as $\bf{q}$, and $\bf{Q}$ is a symmetric tensor. The system is assumed to have the following entropy:
\begin{equation}\label{2.1}
  s=s(u,{\bf{w}},{\bf{Q}})=s^{eq}(u)+s^{neq}(\bf{w},\bf{Q}).
\end{equation}
Here $s^{eq}$ and $s^{neq}$ are both concave functions and correspond to the respective entropy equilibrium part and non-equilibrium part. Based on the generalized Gibbs relations \cite{De2013,jou1996extended}, we deduce the evolution of the entropy:
\begin{equation}\label{2.2}
	\begin{aligned}
		\partial_t s =& s_u \partial_t u + s_{\bf{w}} \cdot\partial_t {\bf{w}} + s_{\bf{Q}} :\partial_t {\bf{Q}}\\
		=&- \theta^{-1} \nabla \cdot {\bf{q}} + {\bf{q}}\cdot \partial_t {\bf{w}} + s_{\bf{Q}}: \partial_t {\bf{Q}}\\
		=&-\nabla\cdot(\theta^{-1} {\bf{q}}+\gamma s_{\bf{Q}} \cdot{\bf{q}}) + \\
         &+{\bf{q}}\cdot(\nabla \theta^{-1}+\gamma \nabla\cdot s_{\bf{Q}} +\partial_t {\bf{w}})+s_{\bf{Q}}:(\gamma \nabla {\bf{q}}+\partial_t {\bf{Q}})\\
         =&-\nabla\cdot(\theta^{-1} {\bf{q}}+\gamma s_{\bf{Q}} \cdot{\bf{q}}) + \\
         &+{\bf{q}}\cdot(\nabla \theta^{-1}+\gamma \nabla\cdot s_{\bf{Q}} +\partial_t \bf{w})\\
         &+s_{\overset{\circ}{\bf{Q}}}:(\gamma {\overset{\circ}{\nabla {\bf{q}}}}+\partial_t {\overset{\circ}{\bf{Q}}})+s_{\overset{\cdot}{Q}}:(\frac{\gamma }{N}\nabla\cdot {\bf{q}}+\partial_t {\overset{\cdot}{Q}})\\
		=:&-\nabla\cdot {\bf{J}} + \sigma.
	\end{aligned}
\end{equation}
Here $\theta^{-1}:=s_u$, and $\gamma$ is a positive constant, and CDF suggests that ${\bf{q}}=s_{\bf{w}}$.
The tensor is decomposed as
${\bf{Q}}=\overset{\circ}{\bf{Q}}+ \overset{\cdot}{ Q}{\bf{I}}$ with $tr ({\overset{\circ}{\bf{Q}}})=0$, and ${\overset{\circ}{\nabla {\bf{q}}}}$ is symmetric where $tr({\overset{\circ}{\nabla {\bf{q}}}})=0$.
${\bf{J}}=\theta^{-1} {\bf{q}}+\gamma s_{\bf{Q}}\cdot{\bf{q}}$ is the entropy flux and $\sigma={\bf{q}}\cdot(\nabla \theta^{-1}+\gamma\nabla \cdot s_{\bf{Q}} +\partial_t {\bf{w}})+s_{\overset{\circ}{\bf{Q}}}:(\gamma {\overset{\circ}{\nabla {\bf{q}}}}+\partial_t {\overset{\circ}{\bf{Q}}})+s_{\overset{\cdot}{Q}}:(\frac{\gamma }{N}\nabla\cdot {\bf{q}}+\partial_t {\overset{\cdot}{Q}})$ is the corresponding entropy production. 

For simplicity, we only consider quasi-one dimensional case in this paper {\color{black}{so that many of vectors mentioned above can be approximated as scalars}}. Since the entropy flux $\sigma$ is nonnegative according to the second law of thermodynamics, we follow CDF \cite{Zhu2015,Yong2020} to obtain that
\begin{subequations}\label{2.3}
\begin{align}
    \partial_t u + \partial_x q =& 0, \label{2.3a}\\
    \partial_t {w} + \partial_x \theta^{-1}+\gamma\partial_x s_{Q}=&{M}_0q ,\label{2.3b}\\
    \partial_t {Q} + \gamma\partial_x {q}=&{M}_1s_{Q}.\label{2.3c}
\end{align}
\end{subequations}
Here ${M}_0={M}_0(u,{w},{Q})$ and ${M}_1={M}_1(u,{w},{Q})$ are two positive functions since $\sigma$ is nonnegative according to the second law of thermodynamics.

\subsection{Specific Model and Discrete Version}
A specific form of the entropy is considered in this paper, that is, the entropy of non-equilibrium part has a quadratic form of the dissipative variables ${w},{Q}$:
\begin{equation}\label{2.4}
  s^{neq}(w,Q)=-\frac{{w}^2}{2\alpha}-\frac{{Q}^2}{2\beta}
\end{equation}
with $\beta$ is a positive constant, and $\alpha=\alpha(u)$ is positive. Thus, \eqref{2.3} becomes
\begin{subequations}\label{2.5}
\begin{align}
    \partial_t u + \partial_x q =& 0 \label{2.5a}\\
    \partial_t (\alpha q) - \partial_x \theta^{-1}+\frac{\gamma}{\beta}\partial_x Q=&-M_0q,\label{2.5b}\\
    \partial_t Q + \gamma\partial_x q=&-\frac{M_1}{\beta}Q\label{2.5c}.
\end{align}
\end{subequations}
Here we have used that $w=-\alpha q$ (i.e., $q=s_w$) and thus, ${M}_0={M}_0(u,{q},{Q})$ and ${M}_1={M}_1(u,{q},{Q})$. Noticing \eqref{2.1} and $\theta^{-1}:=s_u=s^{eq}_u(u)$, we can assert that \eqref{2.5} is well-defined with three unknowns, $u, q$ and $Q$.

Due to the discreteness of the training data (data exist only at discrete space-time points), an alternative discrete version of \eqref{2.5} should be taken into account:
\begin{equation}\label{2.6}
\begin{aligned}
     \frac{u_i^{n+1}-u_i^n}{\Delta t}+\frac{q^n_{i+1}-q^n_{i-1}}{2\Delta x}=&0\\
     (\alpha)^n_i\frac{(q)_i^{n+1}-( q)_i^{n}}{\Delta t} - \frac{ (\theta^{-1})_{i+1}^n-(\theta^{-1})_{i-1}^n}{2\Delta x}+\frac{\gamma}{\beta}\frac{ (Q)_{i+1}^n-(Q)_{i-1}^n}{2\Delta x}=&-(M_0q)^n_i,\\
    \frac{ (Q)_i^{n+1}-(Q)_i^{n}}{\Delta t} + \gamma\frac{ (q)_{i+1}^n-(q)_{i-1}^n}{2\Delta x}=&-\frac{1}{\beta}(M_1Q)_i^n,
\end{aligned}
\end{equation}
where $(\cdot)_i^n$ denote the $(\cdot)'s$ value at space-time points $(i\Delta x,n\Delta t)$, $i$ and $n$ are the indexes of discretized cell and time step, respectively.

It is remarkable that \eqref{2.5} satisfies the conservation-dissipation principle and thereby the structure of the system is globally symmetrizable hyperbolic \cite{Yong2008,Yong2020,XHuoThesis} and it can be easily solved by traditional numerical methods. Further, we should point out that \eqref{2.5} is a generalized model since on the one hand, it is only constrained by basic laws of physics, the first and second laws of thermodynamics, and on the other hand, the model is independent of the data which could come from numerical results, experimental  measurements, or real-world monitoring.

\section{Learn the Unknown Functions and Parameters}
\label{sec:learnAI}

In the last section, we derived the model of heat conduction with unknown functions and parameters in quasi-one dimensional systems by using the CDF. In this section we will show how to learn the unknown functions and parameters in \eqref{2.5} or \eqref{2.6} using deep neural networks, and propose an {\color{black}innovative} method, inner-step operation (ISO), to diminish the extra errors caused by the chosen discrete version.
The current idea of macroscopic heat conduction modeling is the biggest innovation of the present paper, that is, first theoretically deriving a macroscopic heat conduction equation with unknown functions or parameters which is valid at any scale, and then using data-driven deep learning methods to train and learn these unknown functions or parameters.

\subsection{Training Methods}
We rewrite the last two equations of \eqref{2.6} in the following abstract form \cite{han2019uniformly}:
\begin{subequations}\label{3.1}
  \begin{align}
    (\alpha)_i^n(q)_i^{n+1}=&(\alpha q)_i^{n} + \frac{{\Delta t}}{2\Delta x} [(\theta^{-1})_{i+1}^n-(\theta^{-1})_{i-1}^n]-\frac{\gamma}{\beta}\frac{{\Delta t}}{2\Delta x} [(Q)_{i+1}^n-(Q)_{i-1}^n]-\Delta t(M_0q)^n_i\notag\\
    =&:\mathcal{S}[\alpha,\beta,\gamma,M_0](V_{i-1}^n,V_i^n,V_{i+1}^n;\Delta t,\Delta x),\label{3.1a}\\
    (Q)_i^{n+1}=&(Q)_i^{n} - \gamma\frac{\Delta t}{2\Delta x}[ (q)_{i+1}^n-(q)_{i-1}^n]-\frac{\Delta t}{\beta}(M_1Q)_i^n\notag\\
    =&: \mathcal{S}[\beta,\gamma,M_1](V_{i-1}^n,V_i^n,V_{i+1}^n;\Delta t,\Delta x).\label{3.1b}
\end{align}
\end{subequations}
Here $V_i^n=((u)_i^n,(q)_i^n,(Q)_i^n)$, and $\mathcal{S}[\alpha,\beta,\gamma,M_0]$ denotes that $[\alpha,\beta,\gamma,M_0]$ need to be learned via machine learning, similarity for $\mathcal{S}[\beta,\gamma,M_1]$. To be specific, $[\alpha, M_0, M_1]$ are approximated by respective neural networks and $[\beta, \gamma]$ are optimized as parameters.
For \eqref{3.1a} and \eqref{3.1b}, the loss function is defined as:
\begin{subequations}\label{3.2}
  \begin{align}
    L^1_1=&\sum\limits_{\text{training data}}|(\alpha)_i^n(q)_i^{n+1}-\mathcal{S}[\alpha,\beta,\gamma,M_0](V_{i-1}^n,V_i^n,V_{i+1}^n;\Delta t,\Delta x)|^2,\label{3.2a}\\
    L^2_1=&\sum\limits_{\text{training data}}|(Q)_i^{n+1}-\mathcal{S}[\beta,\gamma,M_1](V_{i-1}^n,V_i^n,V_{i+1}^n;\Delta t,\Delta x)|^2,\label{3.2b}
\end{align}
\end{subequations}
where the mean squared error (MSE) is used. In addition,
inspired by the ``warm-up" technique \cite{long2018pde1,long2019pde2} used in the training process, we design loss functions consisting of multi-step time information:
\begin{subequations}\label{3.3}
  \begin{align}
    L^1_k=&\sum\limits_{\text{training data}}|(\alpha)_i^n(q)_i^{n+k}-\mathcal{S}[\alpha,\beta,\gamma,M_0](\tilde{V}_{i-1}^{n+k-1},\tilde{V}_i^{n+k-1},\tilde{V}_{i+1}^{n+k-1};\Delta t,\Delta x)|^2,\label{3.3a}\\
    L^2_k=&\sum\limits_{\text{training data}}|(Q)_i^{n+k}-\mathcal{S}[\beta,\gamma,M_1](\tilde{V}_{i-1}^{n+k-1},\tilde{V}_i^{n+k-1},\tilde{V}_{i+1}^{n+k-1};\Delta t,\Delta x)|^2.\label{3.3b}
\end{align}
\end{subequations}
Here $\tilde{~}$ denotes the outputs of the neural networks, that is, 
\begin{align}
\tilde{V}_i^{n+k-1}&=(\tilde{u}_i^{n+k-1},\tilde{q}_{i}^{n+k-1},\tilde{Q}_{i}^{n+k-1}), \\
\tilde{u}_i^{n+k-1}&=\tilde{u}^{n+k-2}_i-\frac{\Delta t}{2\Delta x}(\tilde{q}_{i+1}^{n+k-2}-\tilde{q}_{i-1}^{n+k-2}) ,  \\
\tilde{q}_{i}^{n+k-1}&=\mathcal{S}[\alpha,\beta,\gamma,M_0](\tilde{V}_{i-1}^{n+k-2},\tilde{V}_i^{n+k-2},\tilde{V}_{i+1}^{n+k-2};\Delta t,\Delta x)/\alpha_i^{n-1} ,\\
\tilde{Q}_{i}^{n+k-1}&=\mathcal{S}[\beta,\gamma,M_1](\tilde{V}_{i-1}^{n+k-2},\tilde{V}_i^{n+k-2},\tilde{V}_{i+1}^{n+k-2};\Delta t,\Delta x) , \\
\tilde{V}_i^{1} &\equiv V_i^1.
\end{align}
Additionally, we also define:
\begin{equation}\label{3.4}
  L^3_k=\sum\limits_{\text{training data}}|u_i^{n+k}-\tilde{u}^{n+k-1}_i-\frac{\Delta t}{2\Delta x}(\tilde{q}_{i+1}^{n+k-1}-\tilde{q}_{i-1}^{n+k-1})|^2.
\end{equation}
Therefore, the total loss function is obtained by combining \eqref{3.3a}, \eqref{3.3b} and \eqref{3.4}:
\begin{equation}\label{3.5}
  L=\sum\limits_{k=1}^K(\lambda_1 L_k^1+\lambda_2 L_k^2+\lambda_3L_k^3),
\end{equation}
where $\lambda_{1,2,3}$ are coefficients.
\subsection{ISO}
To diminish the extra errors caused by the selected discrete version, \eqref{2.6}, we present a method, i.e., the inner-step operation (ISO):
\begin{subequations}\label{3.6}
  \begin{align}
    (\alpha)_i^{n+\frac{k-1}{N}}(q)_i^{n+\frac{k}{N}}=&(\alpha q)_i^{n+\frac{k-1}{N}} + \frac{{\Delta t}}{2N\Delta x} [(\theta^{-1})_{i+1}^{n+\frac{k-1}{N}}-(\theta^{-1})_{i-1}^{n+\frac{k-1}{N}}]-\frac{\gamma}{\beta}\frac{{\Delta t}}{2N\Delta x} [(Q)_{i+1}^{n+\frac{k-1}{N}}-(Q)_{i-1}^{n+\frac{k-1}{N}}]-\frac{\Delta t}{N}(M_0q)^{n+\frac{k-1}{N}}_i,\label{3.6a}\\
    (Q)_i^{n+\frac{k}{N}}=&(Q)_i^{n+\frac{k-1}{N}} - \gamma\frac{\Delta t}{2N\Delta x}[ (q)_{i+1}^{n+\frac{k-1}{N}}-(q)_{i-1}^{n+\frac{k-1}{N}}]-\frac{\Delta t}{N\beta}(M_1Q)_i^{n+\frac{k-1}{N}},\label{3.6b}
\end{align}
\end{subequations}
where $N$ is the inner steps number and $k=1,\cdots,N$. Furthermore, by this method, $\Delta t$ is divided into $\Delta t/N$, and thus it can reduce the dependency between the learned models and the data with extreme small time step ($\Delta t)$.
\subsection{Training Data}
{\color{black}{As mentioned before, the model is independent of the training data which could come from numerical results, experimental  measurements, or real-world monitoring. However, limited by the difficulty of obtaining large amounts of real-world monitoring or experimental data for the present research group,}}  the training data of model in the present paper are obtained from numerically solving the phonon Boltzmann transport equation (BTE) under the Callaway approximation \cite{CZhang2022JAP,PhysRev_callaway,zhang_transient_2021}, which can describe the heat conduction in different regimes~\cite{luo2019,PhysRevB.100.085203,PhysRevB.10.3546,nanoletterchengang_2018,lee_hydrodynamic_2015} and have a good agreement with experimental results \cite{huberman_observation_2019,RevModPhys.90.041002,PhysRevLett.107.095901,PhysRevLett.109.205901},  
\begin{equation}\label{3.7}
\frac{\partial e}{\partial t}+v_g \boldsymbol{s} \cdot \nabla_{\boldsymbol{x}} e=\frac{e_R^{e q}-e}{\tau_R}+\frac{e_N^{e q}-e}{\tau_N}.
\end{equation}
Here $e=e({\boldsymbol{x}}, {\boldsymbol{s}}, t)$ is the phonon distribution function of energy density depending on spatial position ${\boldsymbol{x}}$, unit directional vector ${\boldsymbol{s}}$, time $t$ and group velocity $v_g$. 
$e_R^{e q}$ and $e_N^{e q}$ are the equilibrium state of the momentum-conserved normal scattering process (N-process) and momentum-destroying resistive scattering process (R-process), respectively.
$\tau_R$ and $\tau_N$ are the associated relaxation time, respectively.

\begin{figure}[htb]
  \centering
  \includegraphics[width=0.8\textwidth]{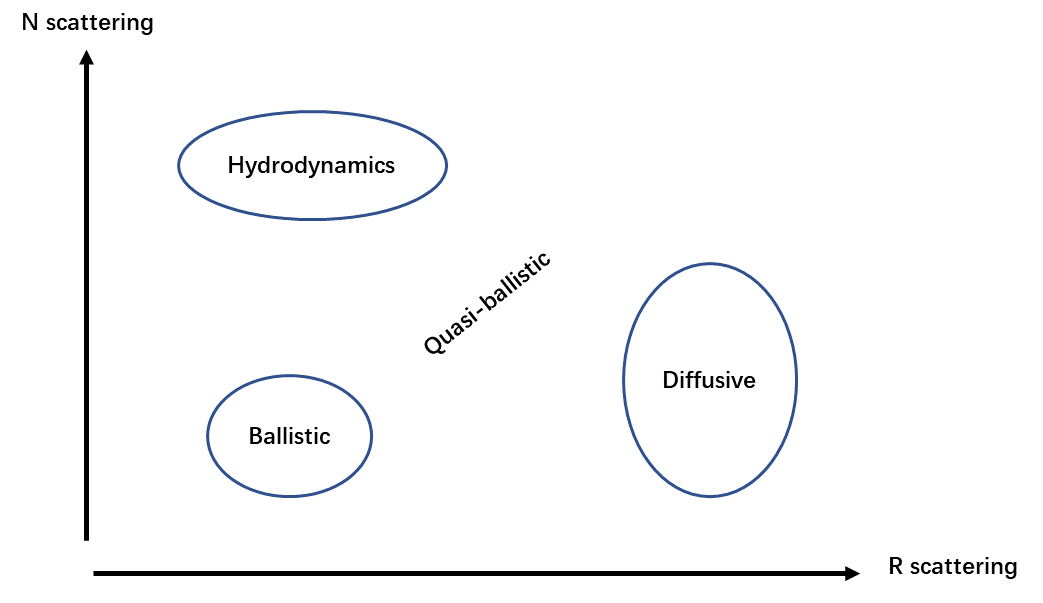}
  \caption{A schematic of phonon transport regimes~\cite{PhysRev_GK,PhysRev.148.766,lee_hydrodynamic_2015,huberman_observation_2019}.}\label{fig_heatwave}
\end{figure}

In this work, the phonon gray model and linear phonon dispersion are used, and the wave vector in three-dimensional materials is assumed to be isotropic. The temperature $\theta$ is assumed that $|\theta-\theta_0|\ll \theta_0$ so that the equilibrium distribution function can be written as follows:
\begin{equation*}
  \begin{aligned}
    e^{eq}_R(\theta)\approx& C\frac{\theta-\theta_0}{4 \pi};\\
    e^{eq}_N(\theta,{\bf{v}})\approx& C\frac{\theta-\theta_0}{4 \pi}+C\theta\frac{\boldsymbol{s}\cdot {\bf{v}}}{4 \pi v_g},
  \end{aligned}
\end{equation*}
where $C=C(\theta_0)$ is the specific heat at reference temperature $\theta_0$, $\bf{v}$ is the drift velocity.
The temperature $\theta$, heat flux $\bf{q}$ and the flux of heat flux $\boldsymbol{Q}$ can be calculated as the moments of distribution functions:
\begin{equation*}
\begin{aligned}
\theta=&\theta_0+\frac{ \int e d \Omega}{ C }, \\
\boldsymbol{q}=&  \int v_g  \boldsymbol{s} e d \Omega,\\
\boldsymbol{Q}=& \int v_g^2 \boldsymbol{ss} e d \Omega,
\end{aligned}
\end{equation*}
where the integral is carried out in the whole solid angle space $d\Omega$.
Here we intend to specify the heat conduction regimes which change with the dimensionless Knudsen numbers defined as
$$
Kn_R^{-1}=\frac{L}{v_g \tau_R},\qquad Kn_N^{-1}=\frac{L}{v_g \tau_N}
$$
which indicates the strength of R- and N-process, respectively, and $L$ is the normalized spatial coordinates. 
Three distinct types of phonon transport are listed below and shown in Fig. \ref{fig_heatwave}:
\begin{itemize}
  \item Hydrodynamics: $Kn_R^{-1}\ll 1 \ll Kn_N^{-1}$;
  \item Ballistic: $Kn_R^{-1}\ll 1$ and $Kn_N^{-1}\ll 1$;
  \item Diffusive: $Kn_R^{-1}\gg 1$.
\end{itemize}

Besides, the Guyer-Krumhansl (G-K) equation can be derived from equation \eqref{3.7} by eigen-value analysis method or Champman-Enskog method \cite{PhysRev_GK,PhysRev.148.766,GUO20151} in the phonon hydrodynamics regime with sufficient N-processes \cite{huberman_observation_2019,CZhang2022JAP,luo2019,nanoletterchengang_2018}, which is a macroscopic heat conduction equation and plays the similar role as Navier-Stokes equation in fluid hydrodynamics. 
The equations reads as:
\begin{equation}\label{3.8}
  \begin{aligned}
    C\partial_t \theta+\nabla \cdot {\bf{q}}=&0,\\
    \tau_R\partial_t {\bf{q}}+{\bf{q}}=&-\kappa\nabla \theta+l^2[\nabla^2 {\bf{q}}+2\nabla(\nabla \cdot {\bf{q}})],
  \end{aligned}
\end{equation}
where the internal energy $u=C\theta$, $\kappa=(1/3)Cv_g^2\tau_R$ is the thermal conductivity and $l^2=(1/5)v_g^2\tau_R\tau_N$.
In this paper, we use dimensionless parameters and set $C=v_g=1$, $\tau_R=0.1,1.0,10$, $\tau_N=0.1,1.0,10$ for pairwise combination.

\subsection{Testing}
\begin{figure}[htb]
  \centering
  \includegraphics[width=0.9\textwidth]{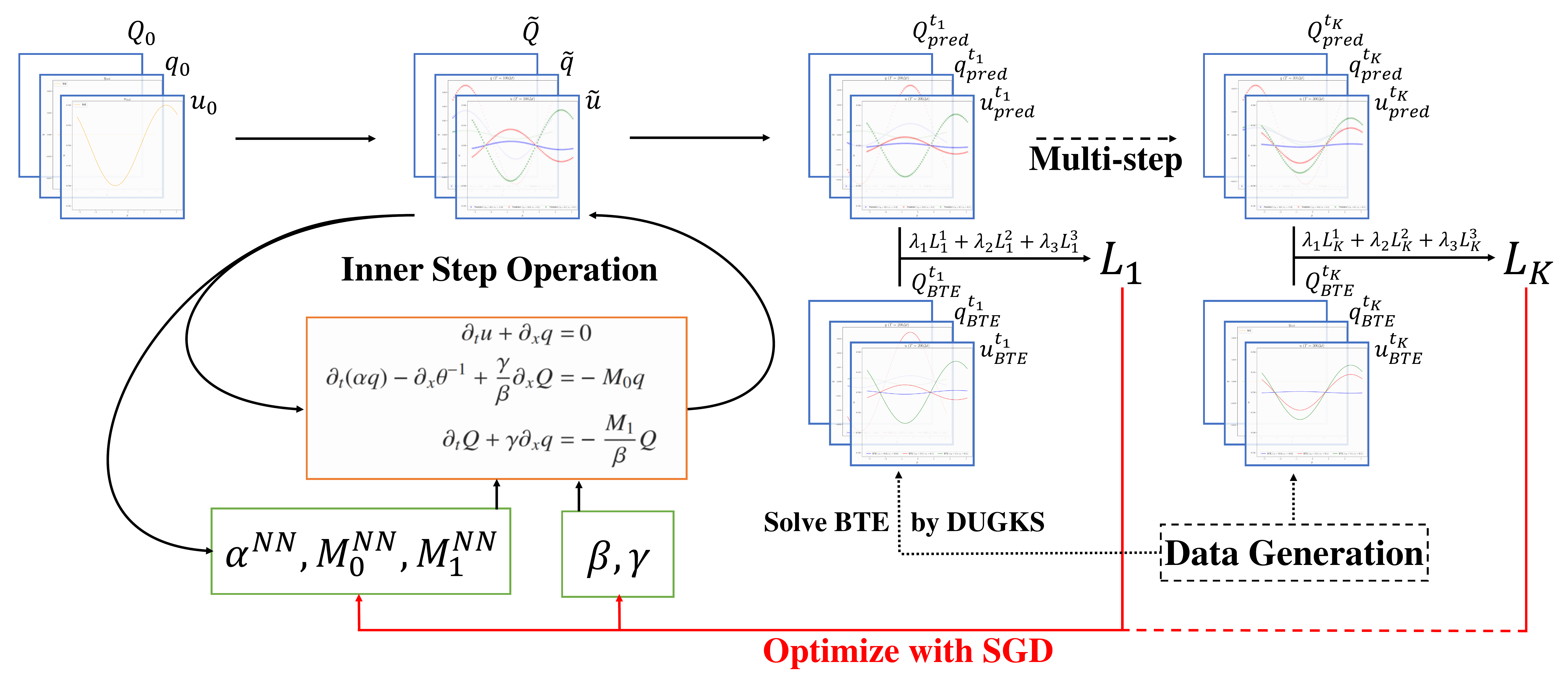}
  \caption{The overview of the framework of the entire training process.}\label{overview_of_framework}
\end{figure}

Once the unknown functions and parameters in the model \eqref{2.5} are trained well, the equations \eqref{2.5} can be solved by traditional numerical methods, such as finite element method, discontinuous Galerkin method and so on. In this paper, we use the finite difference method. The numerical results are validated by the solutions from \eqref{3.7}, which are regarded as ground-truth data. Additionally, we also solve the G-K model \eqref{3.8} numerically in contrast to our model, for purposes of comparison with our model.  We close this section by presenting the overview of the framework of the entire training process in Fig. \ref{overview_of_framework}.

\section{Numerical Results}
\label{sec:numericalresults}

{\color{black}{To show the performance of the present model, the transient heat conduction in quasi-one dimensional system is studied with different N or R scattering rates. 
Similar to the transient thermal grating~\cite{CZhang2022JAP,huberman_observation_2019} or time-domain thermoreflectance experiments~\cite{jiang_tutorial_2018}, an ultrafast heat pulse is implemented on the materials at the initial moment, so that there is a spatial cosine temperature distribution or a hotspot at the center~\cite{zhang_transient_2021}. After the external heat source is removed and the temperature will propagate or dissipate along the $x$ direction.}}

\subsection{Generating Training Data $\&$ Training Setting}
\emph{Generating Training Data.——} We solve the phonon BTE \eqref{3.7} by the discrete unified gas kinetic scheme (DUGKS) \cite{CZhang2022JAP,GuoZl16DUGKS} in the domain $(x,t)\in[-\pi,\pi]\times[0,T]$ with periodic boundary conditions and the initial data constructed below. In this paper, we take $\Delta x=\frac{2\pi}{80}$ and $\Delta t=CFL \times \Delta x$ with $CFL=0.5$. The data are saved every time step, $t=0, \Delta t, 2\Delta t, \cdots, 600\Delta t$. The initial values are constructed as follows:
\begin{equation}\label{4.1}
\begin{aligned}
  \theta(x,t=0)=&\alpha \theta_1(x)+(1-\alpha)\theta_2(x),\\
  \theta_1(x)=&\theta_0+\Delta \theta\cos(x-x_{j_1}),\\
  \theta_2(x)=&\theta_0+\Delta \theta\cos(x-x_{j_2})
\end{aligned}
\end{equation}
with $x_{j_1}=-\pi+{j_1}\times\Delta x$, $x_{j_2}=-\pi+j_2\times\Delta x$, $\theta_0=0.6:0.1:1.0$, $\Delta \theta=0.01:0.02:0.09$ and $\alpha=0:0.2:0.8$.

\emph{Training Setting.——} With respect to the Knudsen number pair, we set three neural networks to approximate $\alpha, M_0, M_1$ in \eqref{3.1}, respectively. All of the neural networks have hidden four layers with fifty neurons in each layer. In order to ensure the positivity of $\alpha, M_0, M_1$, we take \emph{softplus} function as the activation function in the output player, while the \emph{sin} function is used in other layers. We also use adaptive activation functions to accelerate the training \cite{JAGTAP2020109136}. The back propagation (BP) with the stochastic gradient descent (SGD) algorithm is selected as the optimizer and \emph{CosineAnnealingLR} is used to adjust the learning rate \emph{lr}. In the training process, the parameters that need to be optimized is initialized as $\beta=10,~\gamma=1$. For the loss function, we take $K=1$ to ``warm up" the neural networks, and after a number of epochs, we fix $K=4$.

Once the model is learned well, we solve the \eqref{2.5} by the finite difference method \cite{LeVeque1990}.
\subsection{Validation of Accuracy, Long-Term Stability $\&$ Generalization} \label{sec:longtermstability}
\emph{Validation of Accuracy and Long-Term Stability.——} We first show the high accuracy of our learned model \eqref{2.5} and also sketch the solutions of the G-K model \eqref{3.8} for comparison. The initial values are set that $\theta_0=0.75,\Delta \theta=0.01$ in \eqref{4.1} and $q=0$, which is not in the training data. Notice that although the training data are restricted to $[0,600]\Delta t$, we solve the learned model up to $900\Delta t$.

\begin{figure}
\centering
\subfigure[]{
\begin{minipage}[t]{0.45\linewidth}
\centering
\includegraphics[width=2.3in]{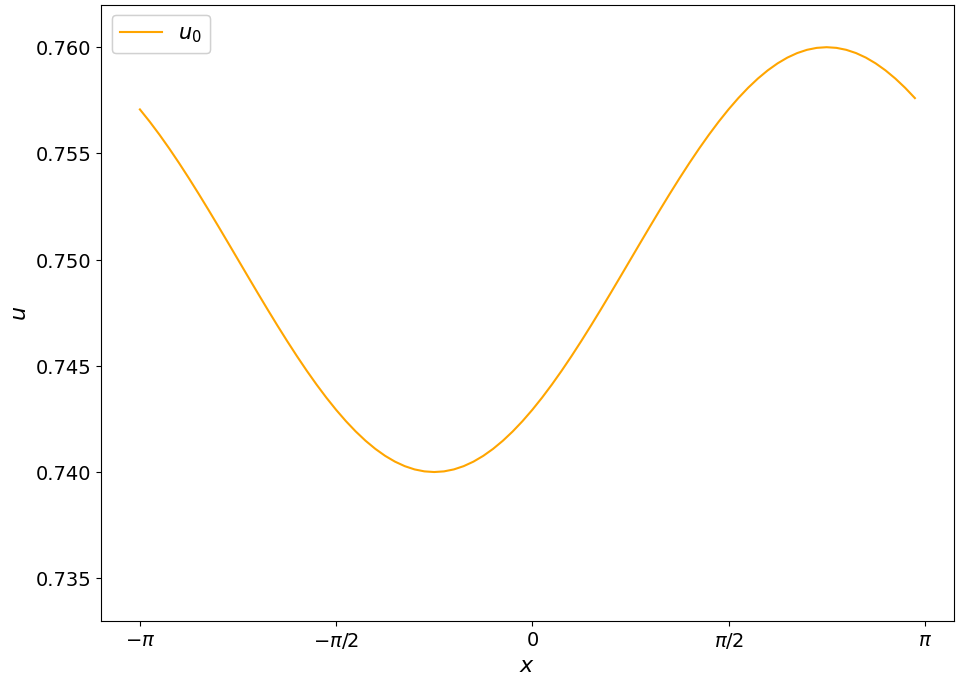}
\end{minipage}%
}%
\subfigure[]{
\begin{minipage}[t]{0.45\linewidth}
\centering
\includegraphics[width=2.3in]{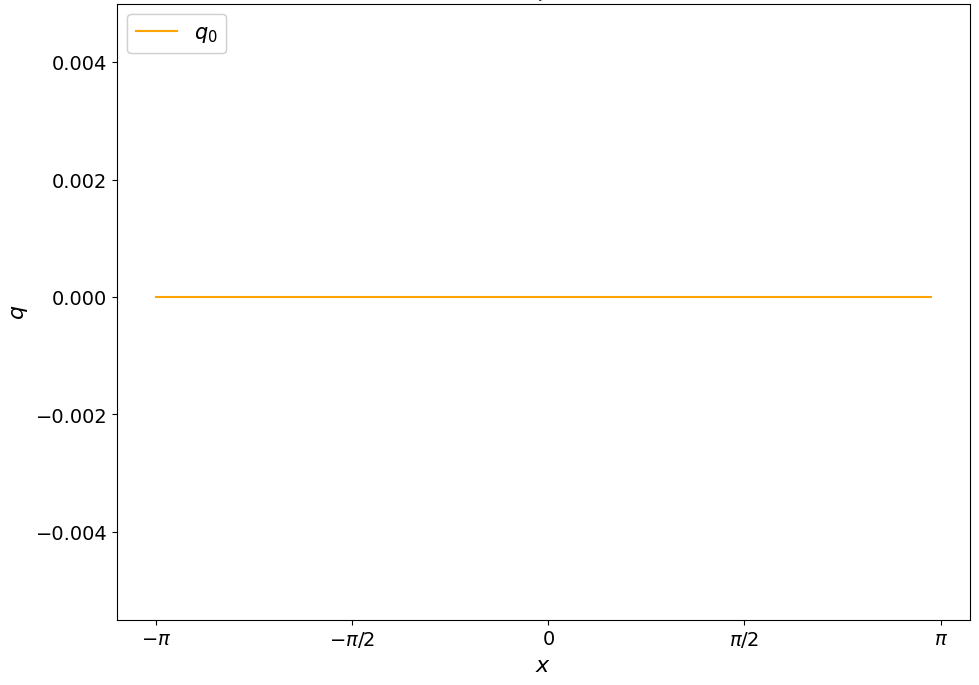}
\end{minipage}%
}%

\subfigure[]{
\begin{minipage}[t]{0.45\linewidth}
\centering
\includegraphics[width=2.3in]{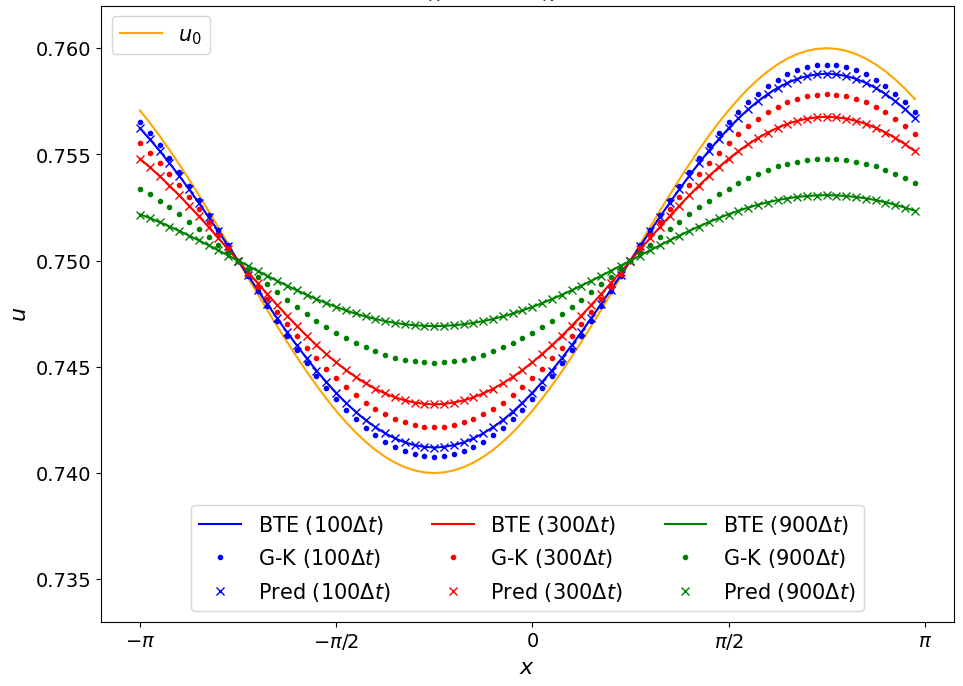}
\end{minipage}
}%
\subfigure[]{
\begin{minipage}[t]{0.45\linewidth}
\centering
\includegraphics[width=2.3in]{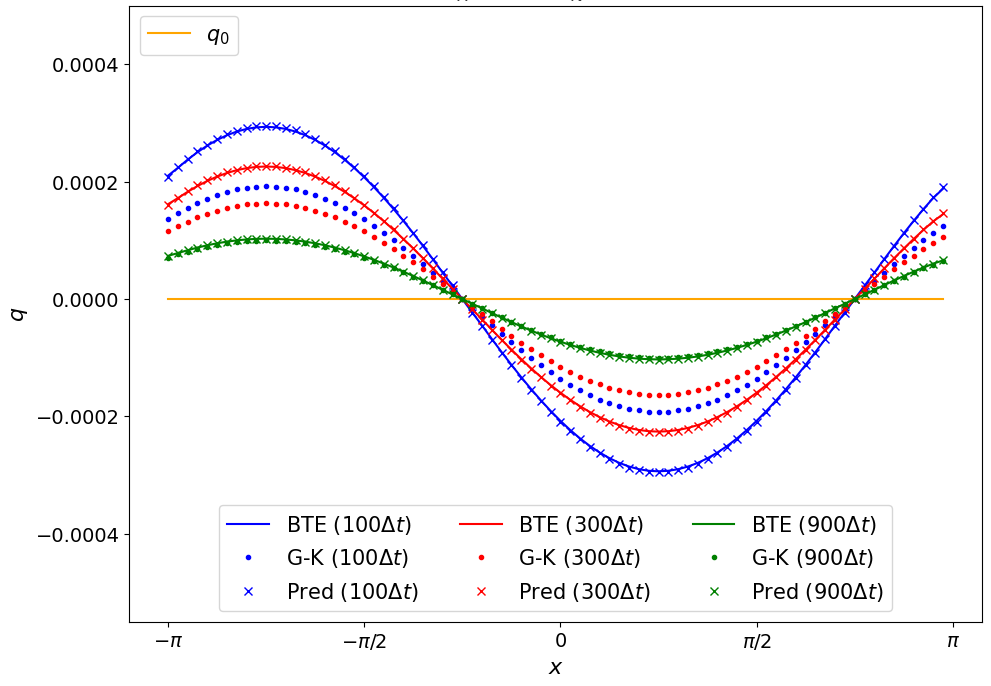}
\end{minipage}
}%

\subfigure[]{
\begin{minipage}[t]{0.45\linewidth}
\centering
\includegraphics[width=2.3in]{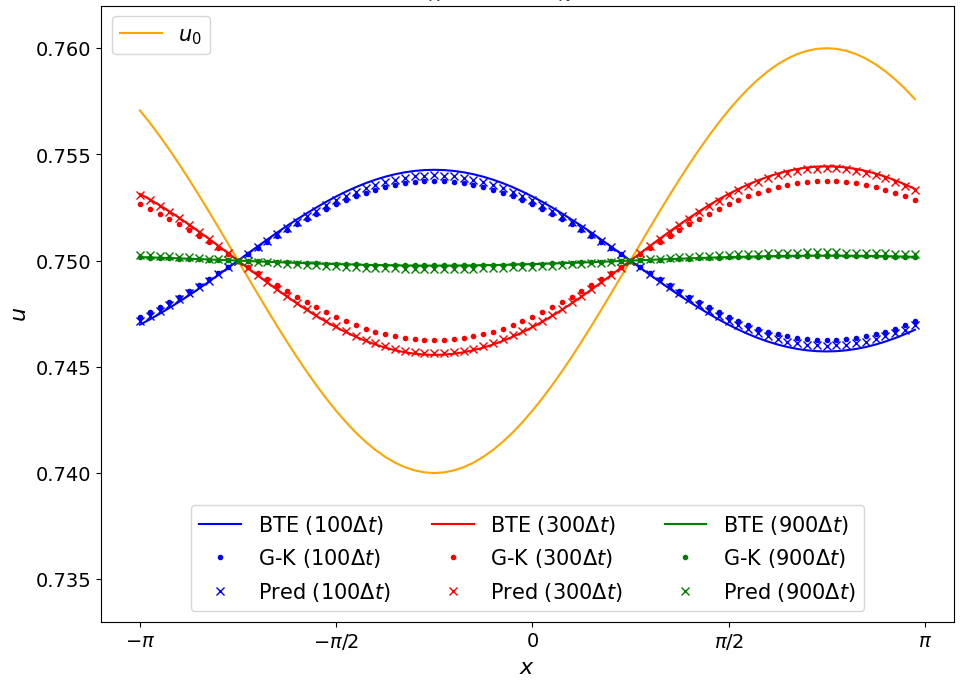}
\end{minipage}
}%
\subfigure[]{
\begin{minipage}[t]{0.45\linewidth}
\centering
\includegraphics[width=2.3in]{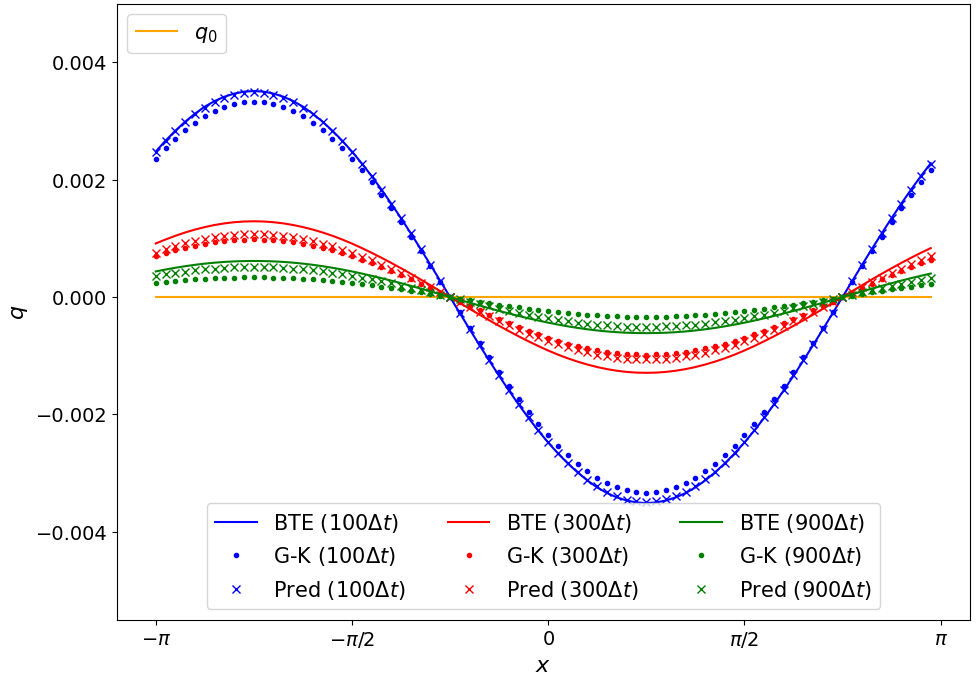}
\end{minipage}
}%

\subfigure[]{
\begin{minipage}[t]{0.45\linewidth}
\centering
\includegraphics[width=2.3in]{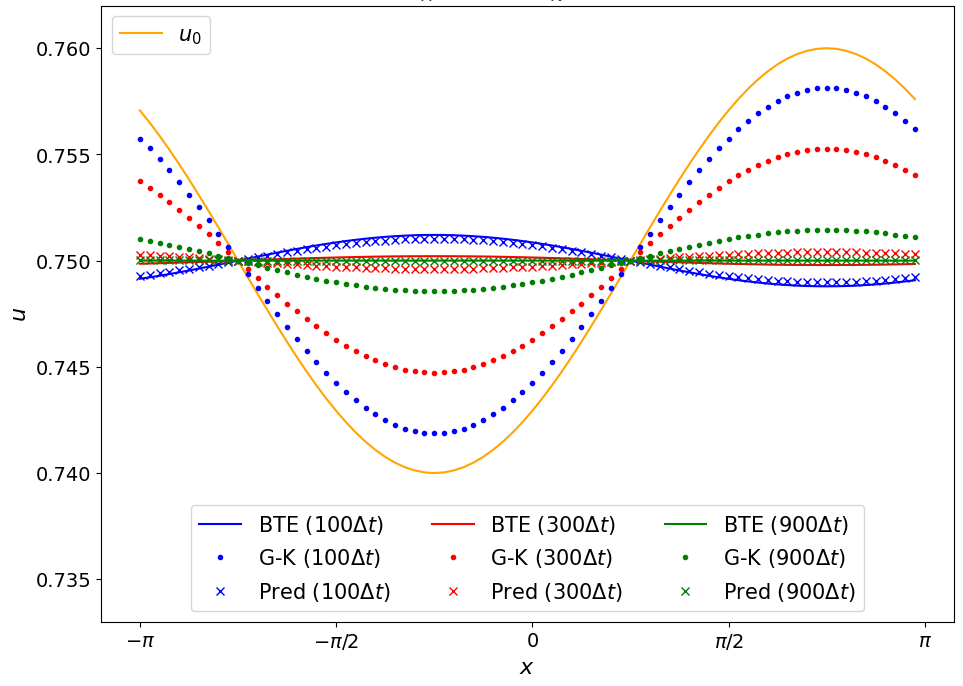}
\end{minipage}
}%
\subfigure[]{
\begin{minipage}[t]{0.45\linewidth}
\centering
\includegraphics[width=2.3in]{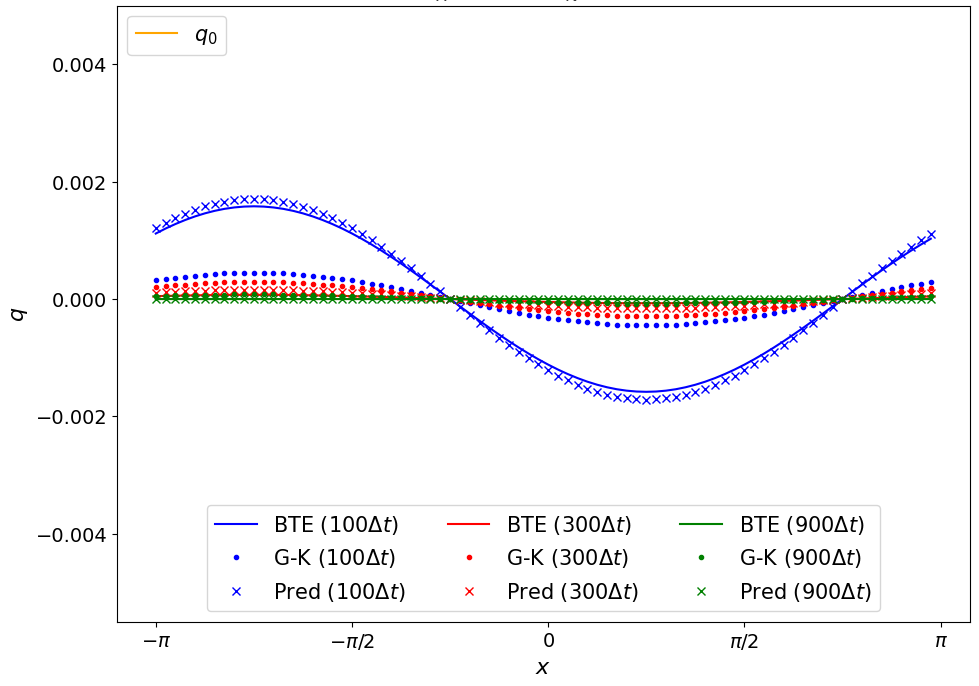}
\end{minipage}
}%
\centering
\caption{The comparison of solutions of our model and the G-K model at $100\Delta t$, $300\Delta t$ and $900\Delta t$. The baseline is the solutions of the BTE. (a,b) The first row is the initial data about $u$ and $q$, respectively. The last three rows are about (c,d) diffusive $\tau_R=0.1,~\tau_N=10$, (e,f) hydrodynamics $\tau_R=10,~\tau_N=0.1$ and (g,h) ballistic cases $\tau_R=10,~\tau_N=10$, respectively.}\label{fig4.2.1}
\end{figure}

The results with respect to diffusive ($\tau_R=0.1,\tau_N=10$), hydrodynamics ($\tau_R=10,\tau_N=0.1$), and ballistic ($\tau_R=10,\tau_N=10$) regimes are plotted in Fig. \ref{fig4.2.1}.
Here the solutions of the BTE are taken as a benchmark, and the snapshots at $100\Delta t$, $300\Delta t$ and $900\Delta t$ are shown. It is easy to see that our model can predict the solutions with high accuracy, and performs better than those of the G-K model.

~~~~\\

\begin{figure}
\centering
\subfigure[]{
\begin{minipage}[t]{0.45\linewidth}
\centering
\includegraphics[width=2.3in]{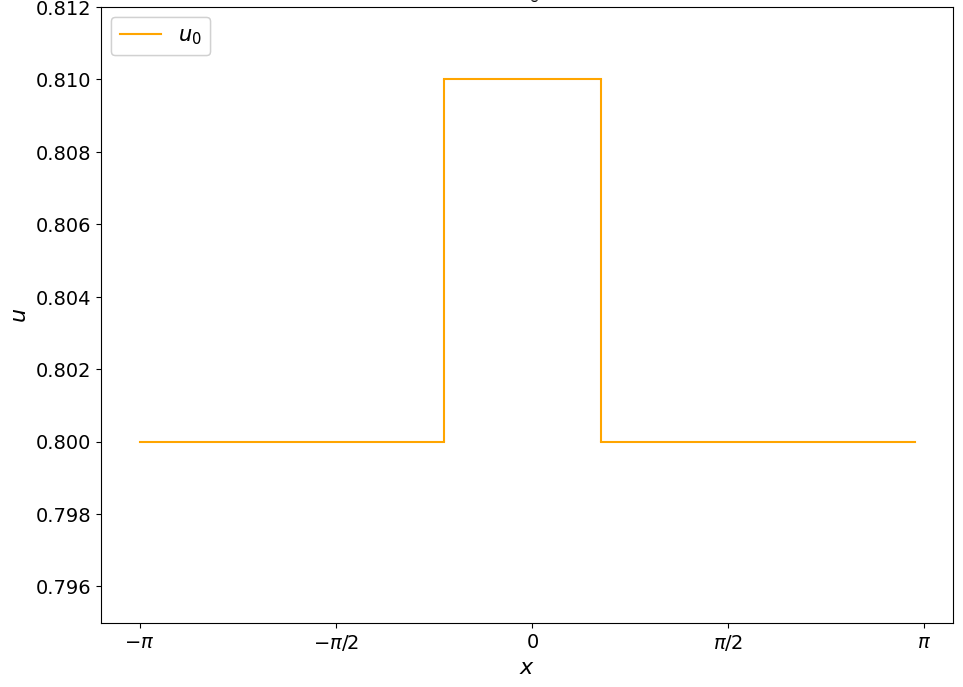}
\end{minipage}%
}%
\subfigure[]{
\begin{minipage}[t]{0.45\linewidth}
\centering
\includegraphics[width=2.3in]{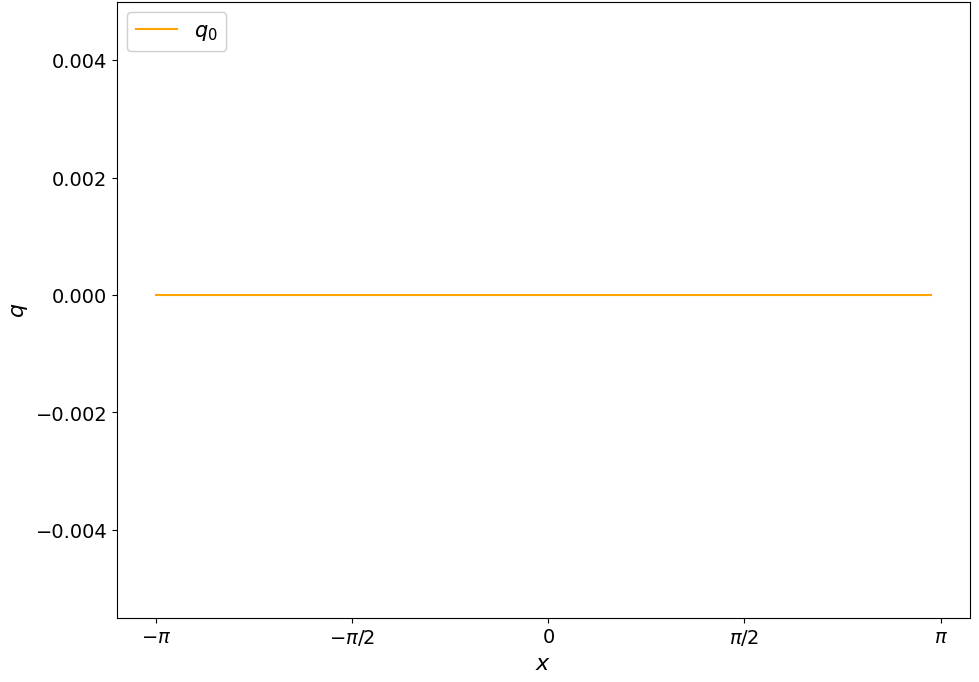}
\end{minipage}%
}%

\subfigure[]{
\begin{minipage}[t]{0.45\linewidth}
\centering
\includegraphics[width=2.3in]{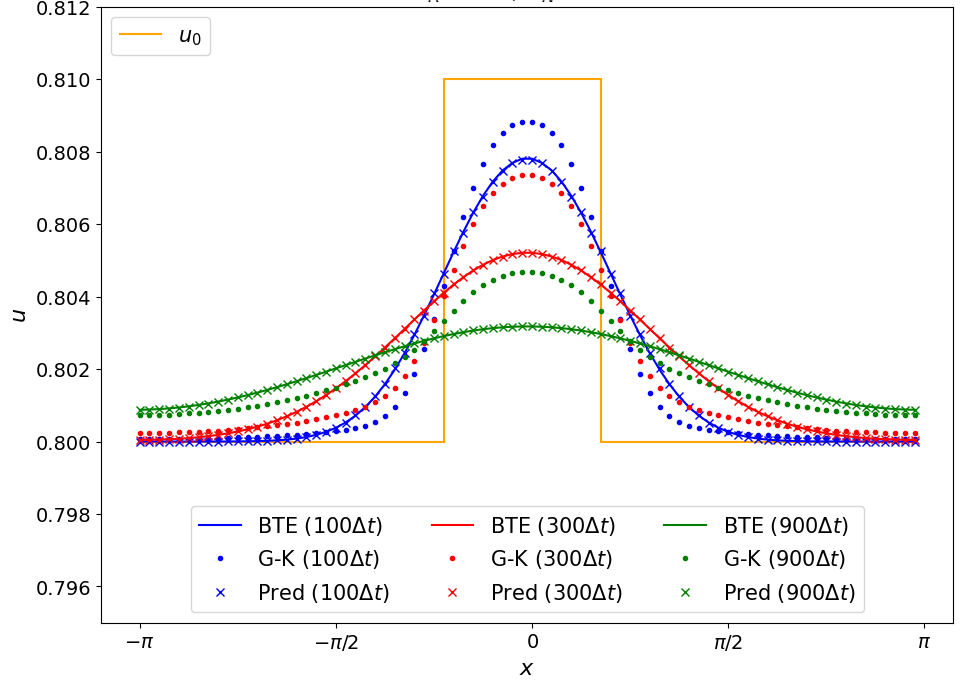}
\end{minipage}
}%
\subfigure[]{
\begin{minipage}[t]{0.45\linewidth}
\centering
\includegraphics[width=2.3in]{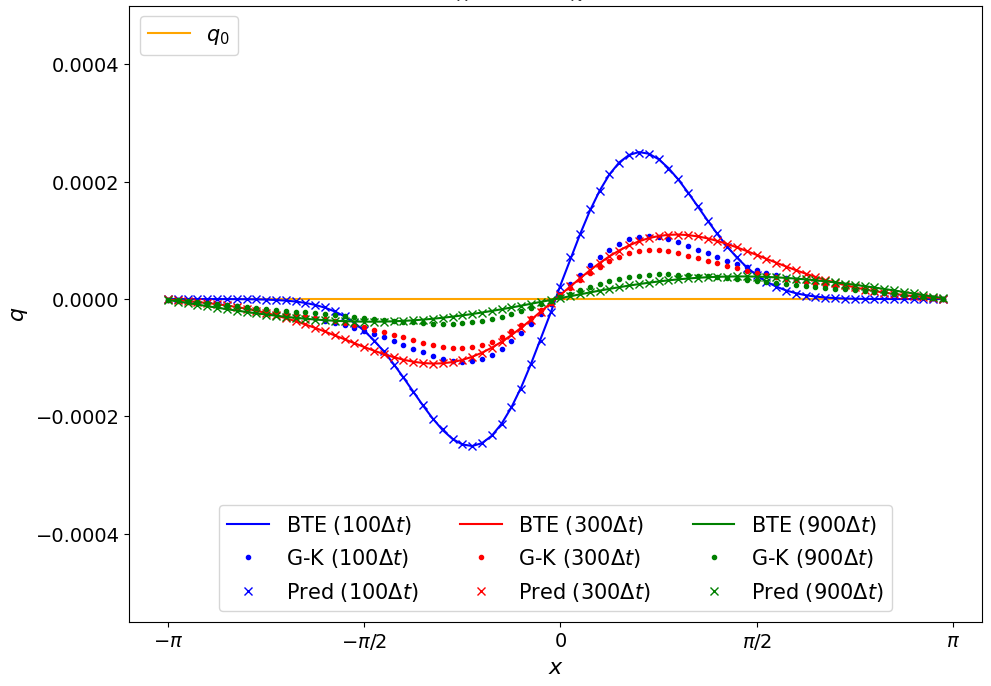}
\end{minipage}
}%

\subfigure[]{
\begin{minipage}[t]{0.45\linewidth}
\centering
\includegraphics[width=2.3in]{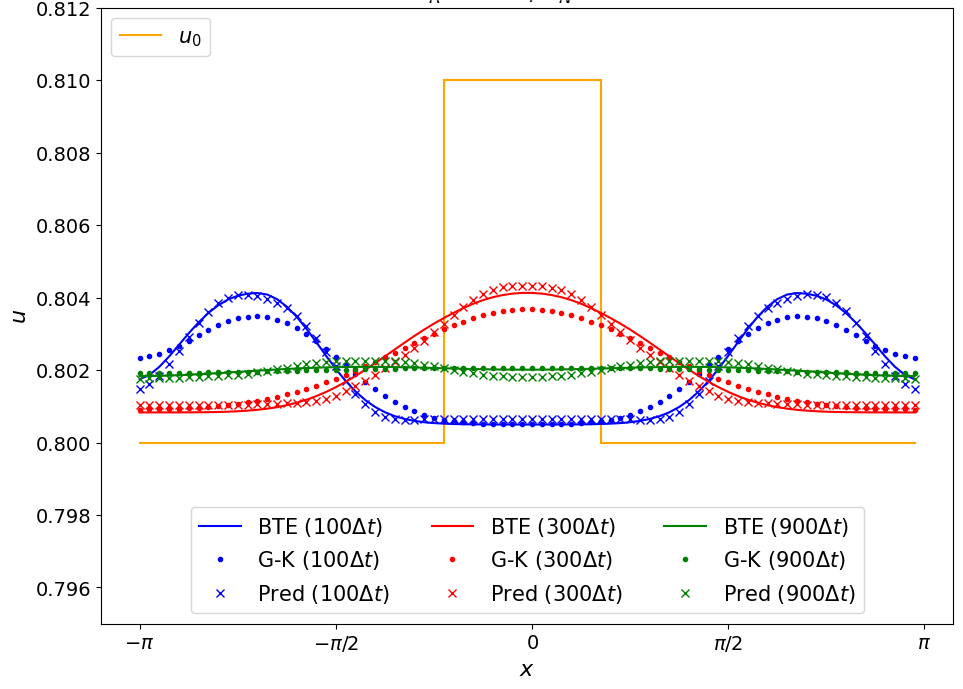}
\end{minipage}
}%
\subfigure[]{
\begin{minipage}[t]{0.45\linewidth}
\centering
\includegraphics[width=2.3in]{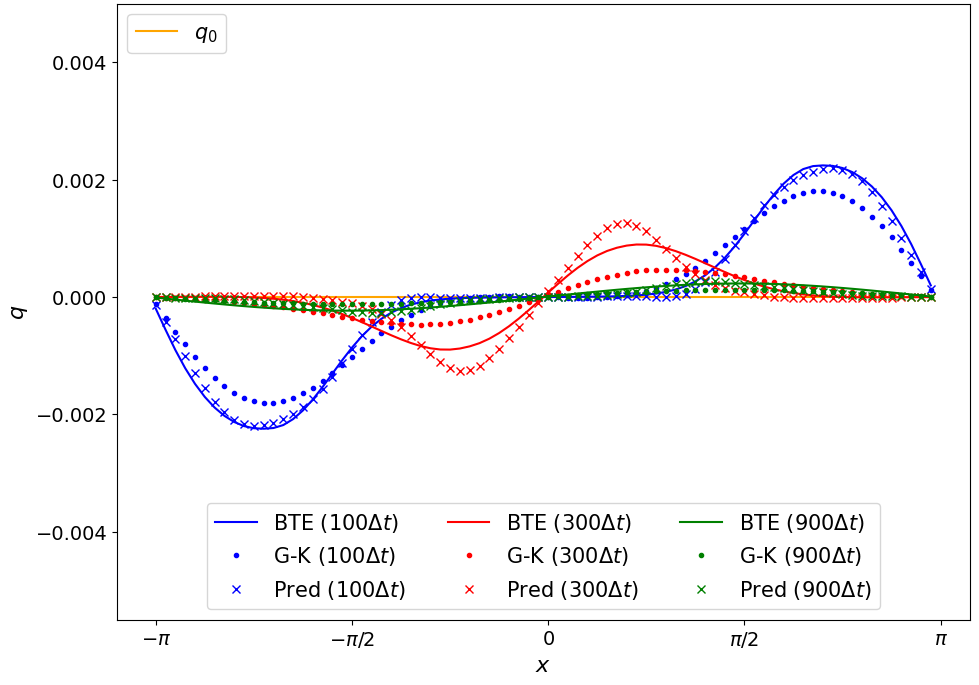}
\end{minipage}
}%

\subfigure[]{
\begin{minipage}[t]{0.45\linewidth}
\centering
\includegraphics[width=2.3in]{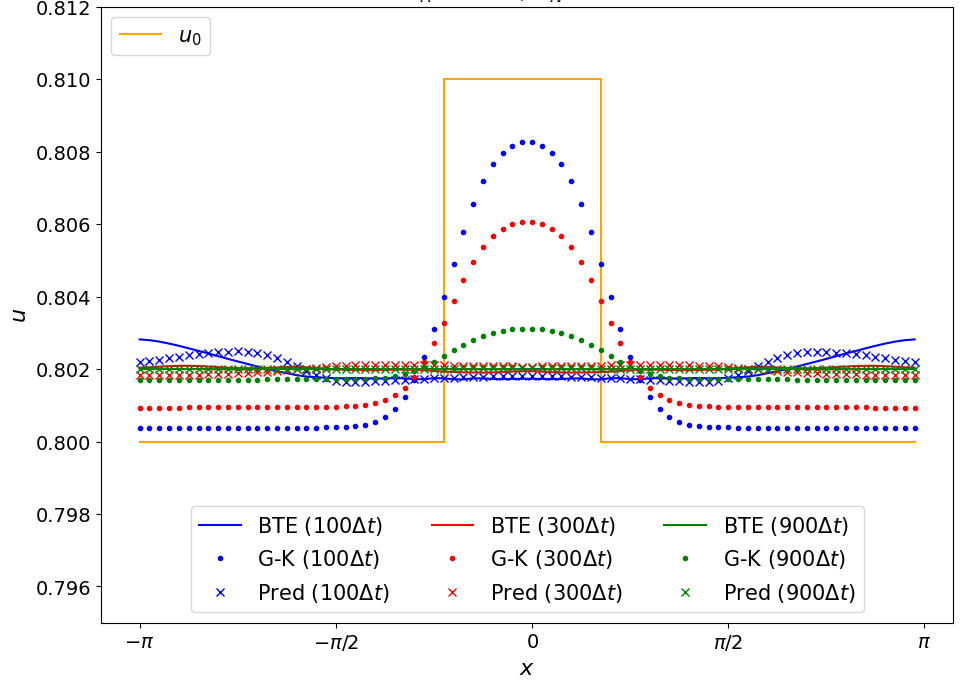}
\end{minipage}
}%
\subfigure[]{
\begin{minipage}[t]{0.45\linewidth}
\centering
\includegraphics[width=2.3in]{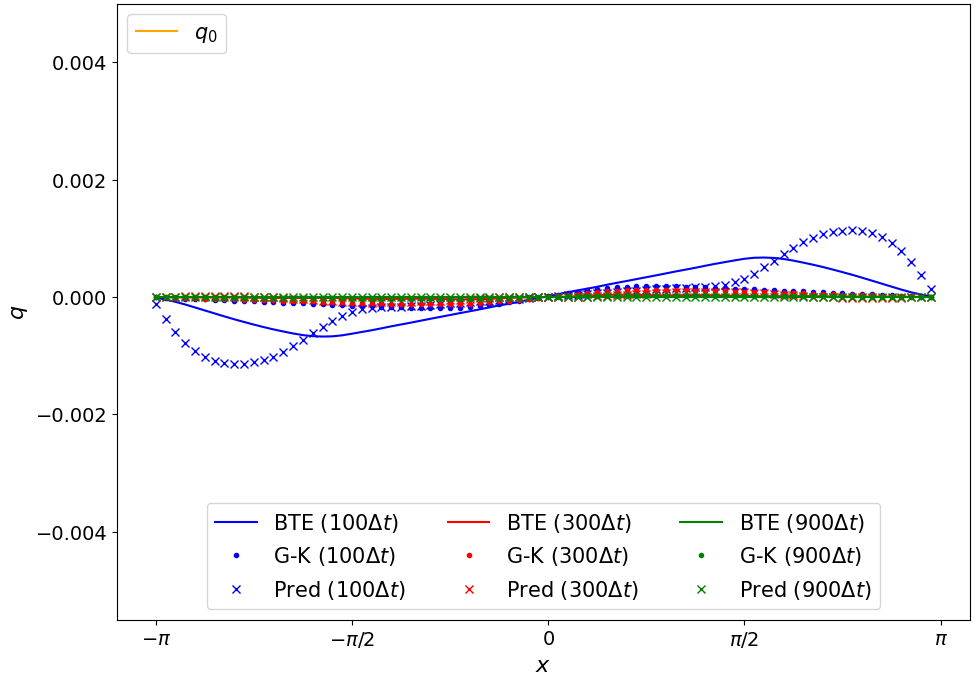}
\end{minipage}
}%
\centering
\caption{The results of our model with discontinuous initial values at $100\Delta t$, $300\Delta t$ and $900\Delta t$. The results of the G-K model are also plotted for comparison, and the baseline is the solutions of the BTE. (a,b) The first row is the initial data about $u$ and $q$, respectively. The last three rows are about (c,d) diffusive $\tau_R=0.1,~\tau_N=10$, (e,f) hydrodynamics $\tau_R=10,~\tau_N=0.1$ and (g,h) ballistic cases $\tau_R=10,~\tau_N=10$, respectively.}\label{fig4.2.2}
\end{figure}

\emph{Validation of Generalization.——} The performance of the learned model with discontinuous initial values is explored in this experiment. The results are demonstrated in Fig \ref{fig4.2.2}, of which the first row is about the initial values of $u$ and $q$. For comparison, we also plot the results of the G-K model. We should point out again that the training data only consists of smooth cases. From Fig \ref{fig4.2.2}, we can say that the results of our model agree with the exact ones very well, especially in the ballistic case, while the G-K model disaccords with the BTE results.
This implies that our model is valid with discontinuous initial data and has good generalization.

\subsection{Heat Waves}   \label{sec:heatwaves}
\begin{figure}[htb]
  \centering
  \includegraphics[width=0.55\linewidth]{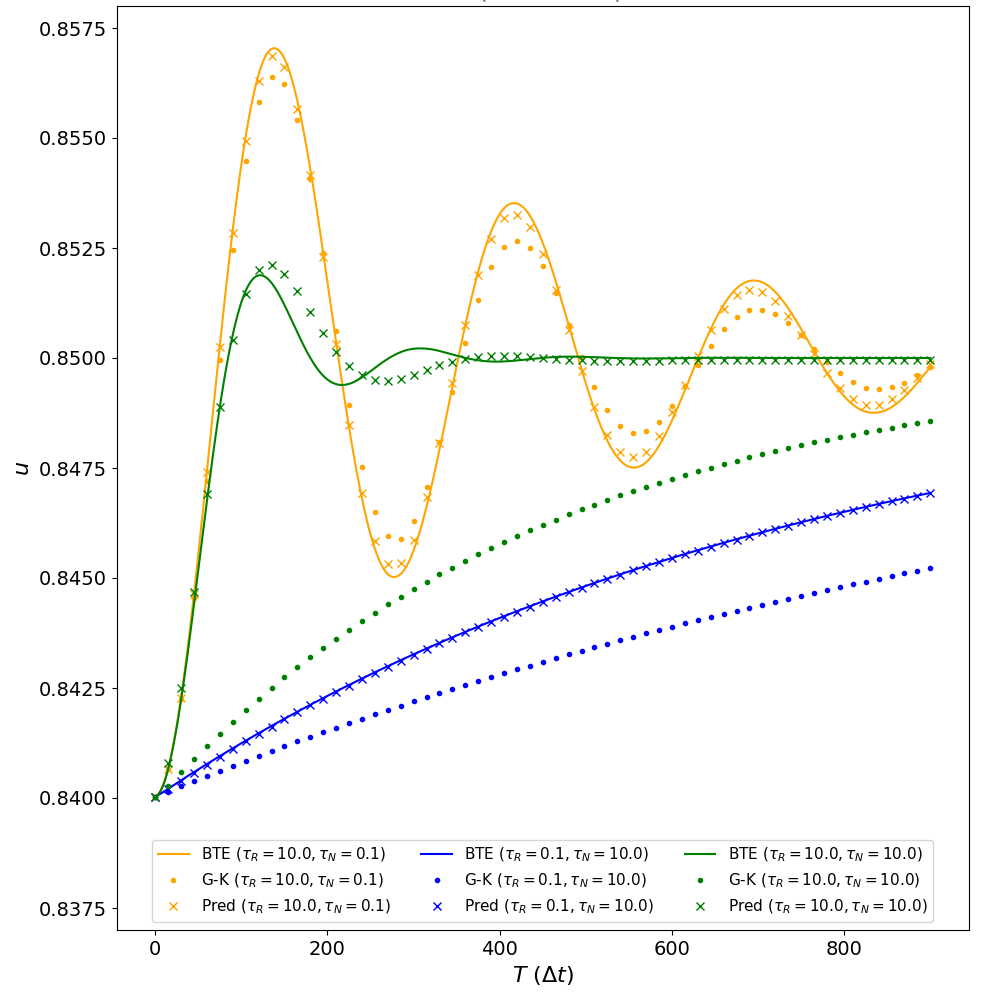}
  \caption{The heat waves of $u(-\pi+30\Delta t)$ with different Knudsen numbers. We take BTE as a baseline, and also show the behaviors of the G-K model.}\label{fig4.3}
\end{figure}

In this subsection, we test the behaviors of predicting the heat waves by the model with dual-dissipative variables, such as diffusive, hydrodynamic and ballistic types, the last two of which can not be captured by the model with one-dissipative variable \cite{ZHAO2022122396}.
The total results are drawn in Fig \ref{fig4.3}.
When $\tau_R=0.1$, the heat conduction is in the diffusive regime, and the internal energy $u$ decays over time; when $\tau_N=0.1,~ \tau_R=10$, the heat conduction is in the hydrodynamic regimes, and the internal energy $u$ decays slowly with large heat waves; when $\tau_N=10,~ \tau_R=10$, the heat conduction is in the ballistic regimes, and the internal energy $u$ decays quickly with small heat waves. From Fig \ref{fig4.3}, it is easy to see that the results of our model are consistent with those of the BTE very well.

\subsection{ISO}
\begin{figure}[htb]
\centering
\subfigure[]{
\begin{minipage}[t]{0.45\linewidth}
\centering
\includegraphics[width=2.8in]{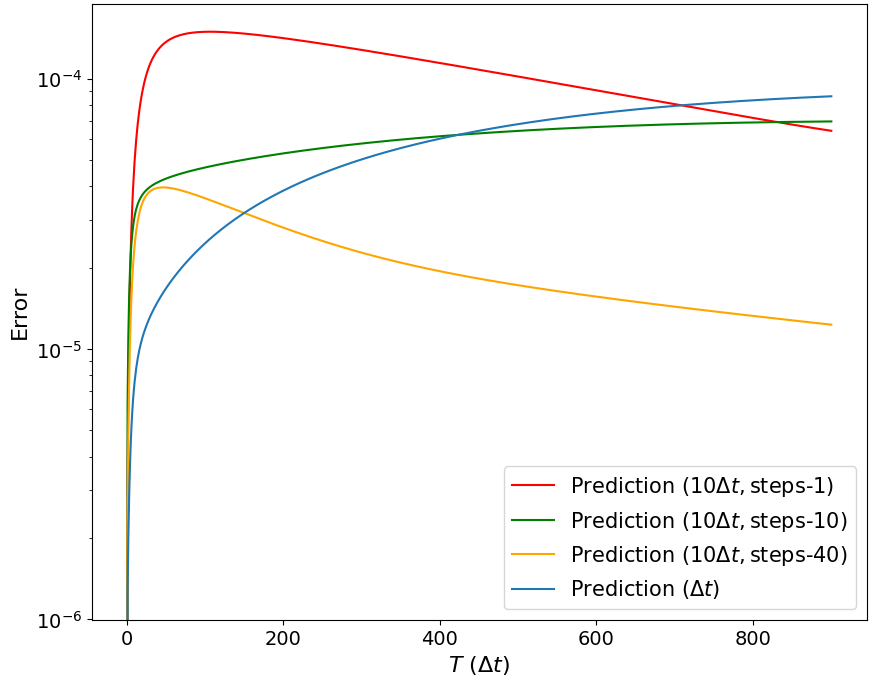}
\end{minipage}%
}%
\subfigure[]{
\begin{minipage}[t]{0.45\linewidth}
\centering
\includegraphics[width=2.8in]{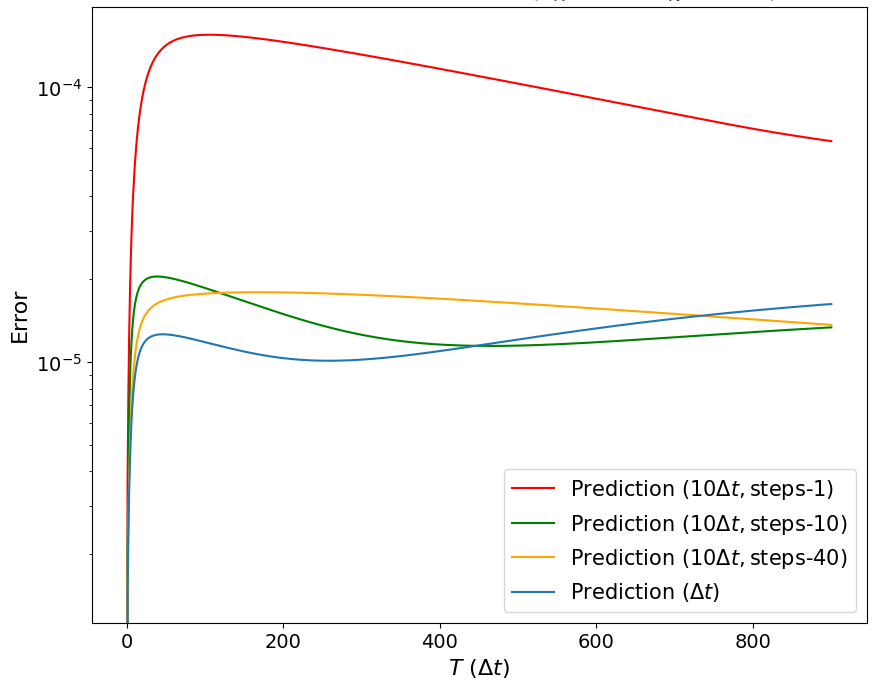}
\end{minipage}%
}%

\subfigure[]{
\begin{minipage}[t]{0.45\linewidth}
\centering
\includegraphics[width=2.8in]{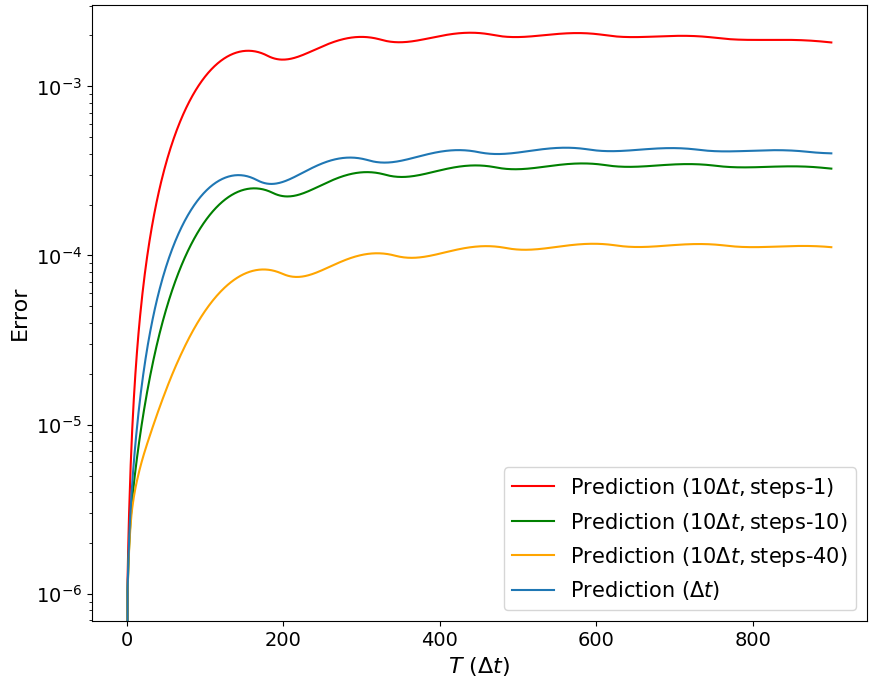}
\end{minipage}
}%
\subfigure[]{
\begin{minipage}[t]{0.45\linewidth}
\centering
\includegraphics[width=2.8in]{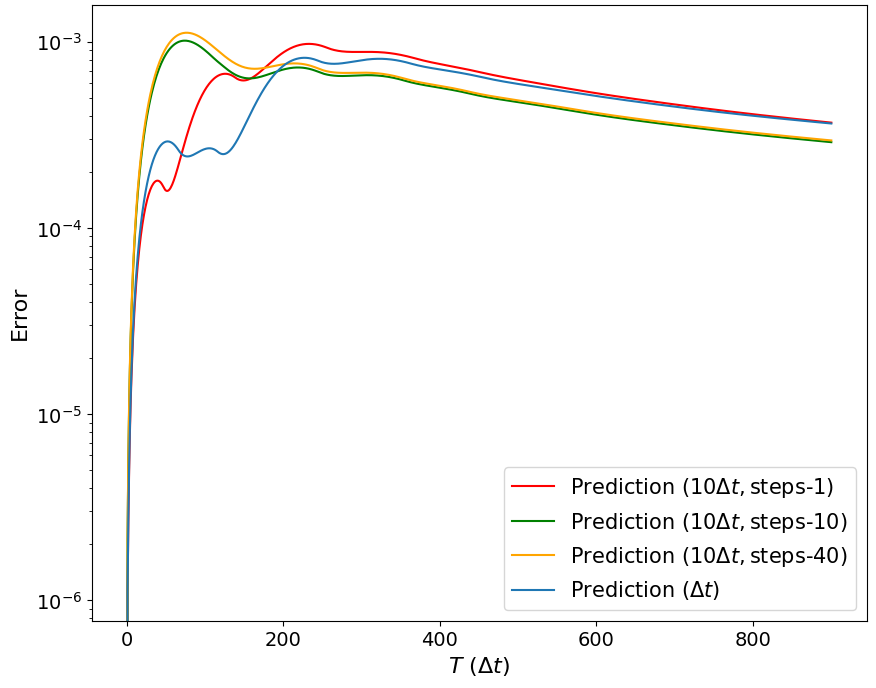}
\end{minipage}
}%
\centering
\caption{The mean$-L^2$ relative errors of $u$ with different Knudsen numbers. Here step$-\ast$ means ISO method with $\ast$ inner steps. Prediction ($\Delta t$) denotes that the model is trained by original data of every time step ($\Delta t$).  (a) $\tau_R=0.1,~\tau_N=0.1$, (b) $\tau_R=0.1,~\tau_N=10$, (c) $\tau_R=10,~\tau_N=0.1$, (d) $\tau_R=10,~\tau_N=10$.  }\label{fig4.4.1}
\end{figure}

In this section we will demonstrate the power of the inner-step operation (ISO) in the mechanism-data fusion modeling process. The ISO is {\color{black}advanced} and plays an important role in diminishing the extra errors caused by the alternative discrete version. For this purpose, we conduct the following experiment: we only choose $60$ points uniformly in $[0,T]$ to train the model, not the original $600$ points, i.e., $t=0, 10\Delta t, 20\Delta t, \cdots, 600\Delta t$. In the training process, we use the ISO with inner step= $1,~10,~40$, and we also show the original results for comparison. Here we define the mean$-L^2$ relative error as
$$
L=\frac{1}{N}\sum\limits_{i=1}^N\frac{\sqrt{\sum\limits_{j=1}^{80}(u-u_{BTE})^2(x_j,t_i)}}{\sqrt{\sum\limits_{j=1}^{80}u_{BTE}^2(x_j,t_i)}},\qquad N=1,2,\cdots,900.
$$
The corresponding results are demonstrated in Fig \ref{fig4.4.1}, which shows that the model trained by coarse data with the ISO method can be competitive with that trained with fine data. Besides, the ISO method performs better with larger inner steps.

\begin{figure}[htb]
\centering
\subfigure[]{
\begin{minipage}[t]{0.45\linewidth}
\centering
\includegraphics[width=2.8in]{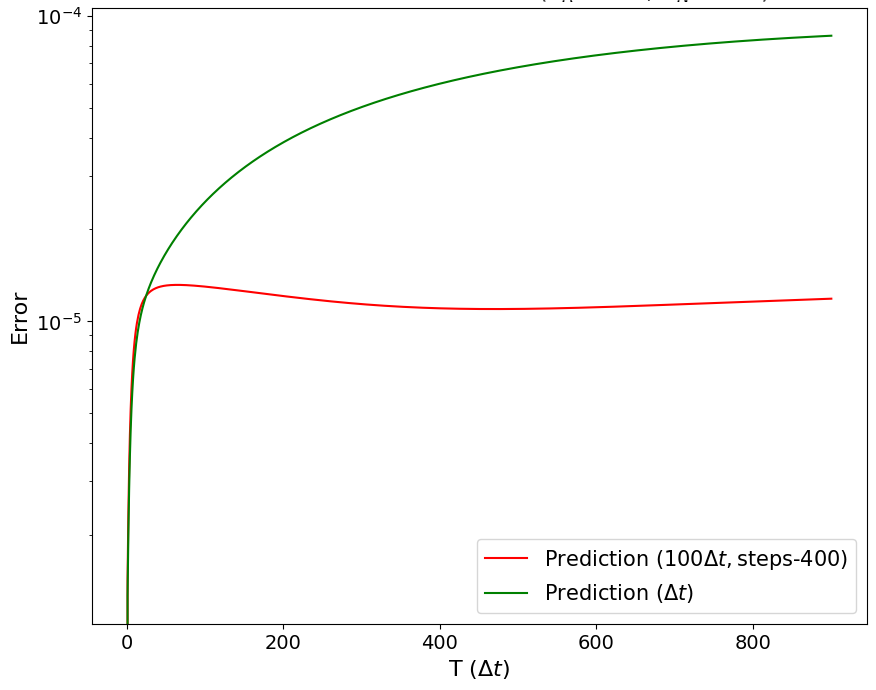}
\end{minipage}%
}%
\subfigure[]{
\begin{minipage}[t]{0.45\linewidth}
\centering
\includegraphics[width=2.8in]{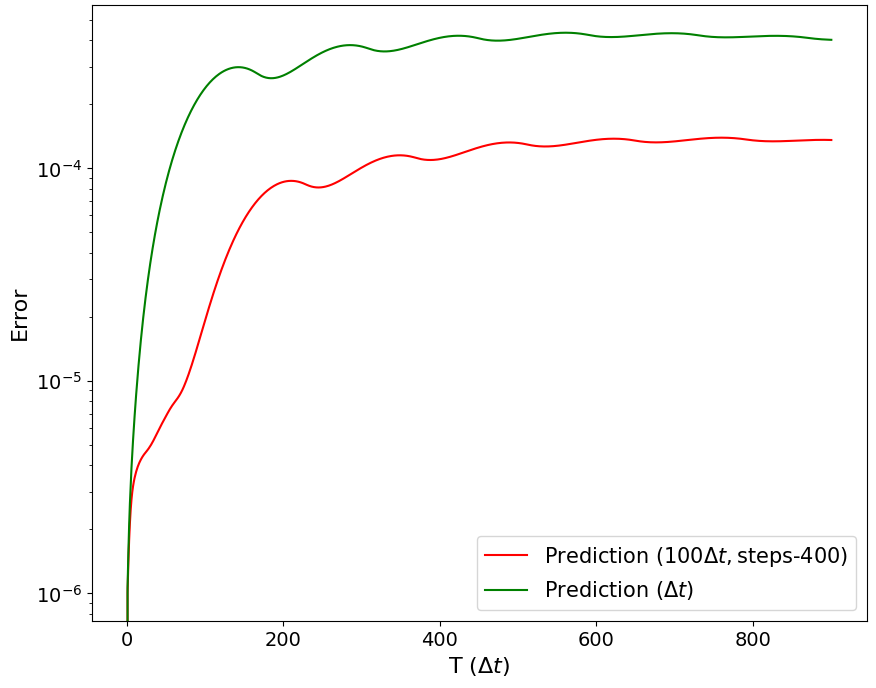}
\end{minipage}%
}%
\centering
\caption{The mean$-L^2$ relative errors of $u$ with the ISO method. Here step$-\ast$ means ISO method with $\ast$ inner steps. Prediction ($\Delta t$) denotes that the model is trained by original data of every time step ($\Delta t$). (a) $\tau_R=0.1,~\tau_N=0.1$ and (b) $\tau_R=10.0,~\tau_N=0.1$  }\label{fig4.4.2}
\end{figure}

Moreover, we test the case that the data consist of coarser time steps, $t=0, 100\Delta t, 200\Delta t, \cdots, 600\Delta t$, i.e., 6 points in $[0,T]$. In this situation, the model can not be trained well without the ISO method. However, with the ISO method, the model still has excellent performances. From Fig \ref{fig4.4.2}, it is obvious that the ISO method plays a vital role in the model training process, and the prediction with the ISO method is even better than the original one.

\section{Analysis and discussions}
\label{sec:analysis}

In this section, we summarize, analyze and discuss the proposed model and training methods.
Different from most of previous macroscopic non-Fourier heat conduction models obtained by Chapman-Enskog, Hermite or other small perturbation expansion methods, the heat conduction model built by the present framework is not limited by small Knudsen numbers.
The current framework of macroscopic heat conduction modeling is the biggest innovation of the present paper, that is, first theoretically deriving a macroscopic heat conduction equation with unknown functions or parameters which is valid at any scale, and then using data-driven deep learning methods to train and learn these unknown functions or parameters.

We advocate for a {\color{black}groundbreaking} Mechanism-Data Fusion Method (MDFM), designed to model heat conduction using a dual-dissipative variables approach. The meteoric rise in computational processing capabilities and exponential growth in data storage capacities present opportunities for synergizing mechanism-based modeling with data-driven discoveries of nascent physical principles \cite{brunton2016discovering,raissi2019physics,Edalatifar2021,WOS:000866513700001,WOS:000971515700001,htj.22255}.  
Advances in statistical methodologies coupled with machine learning's evolution have considerably enhanced the potency of data-driven models. The ascendancy of these computational tools, such as deep neural networks (DNN), recurrent neural networks (RNN), and convolutional neural networks (CNN), have been notable. However, that most machine-learning-based models lack explicit expressions creates a barrier for the interpretability of the physical models owing to the reputation of being referred to as ``black box".

We amalgamate machine learning methodologies with the Conservation-Dissipation Formalism (CDF) \cite{Zhu2015,Yong2020}, thus deriving an explicit and interpretable series of partial differential equations (PDEs) that elucidate both Fourier and non-Fourier modes of heat conduction. Utilizing the CDF, we introduce dual-dissipative variables to formulate a system characterized by first-order hyperbolic PDEs. Anchored in the second law of thermodynamics, the CDF ensures that the resultant PDE system adheres to the conservation-dissipation principle \cite{Yong2008} and manifests as a globally symmetrizable hyperbolic structure \cite{XHuoThesis}. Subsequent to this theoretical foundation, we employ deep neural networks (DNNs) to train the unknown functions within these PDEs, utilizing a preliminary ``warm-up" procedure \cite{long2018pde1} to streamline the integration of temporal data series. Notably, while the data utilized for training in this study is derived from the phonon Boltzmann transport equation (BTE) \cite{CZhang2022JAP}, the training paradigm itself remains indifferent to the PDE derivation process, thereby accommodating data from a spectrum of sources, including numerical analyses, empirical simulations, or direct empirical observation.

In the training procedure, we propose the inner-step operation (ISO), an ingenious strategy engineered to bridge the gap between discrete formulations and the continuous paradigm inherent in system modeling. The discrete nature of our training data necessitates the adoption of an approximate discrete representation of the governing PDEs, which invariably introduces numerical discrepancies. Given DNN's insensitivity to higher-order discretizations, which paradoxically may induce instabilities, our ISO methodology seeks to attenuate the extraneous errors engendered by the choice of discretization schemes. Empirical evidence amassed through a plethora of numerical tests corroborates our assertion that the ISO methodology not only significantly mitigates these errors but also lessens the model's sensitivity to data characterized by infinitesimal time-step magnitudes.

Lots of numerical tests are carried out to show the performance of the proposed model. It bears mentioning that the learned model is amenable to resolution via traditional numerical techniques, such as the Finite Element Method (FEM), Finite Difference Method (FDM), Finite Volume Method (FVM), alongside more contemporaneous approaches like the Deep Ritz Method (DRM) \cite{DRM2018}, Deep Galerkin Method (DGM) \cite{SIRIGNANO20181339}, and Physics-Informed Neural Networks (PINN) \cite{raissi2019physics}. An attribute of our endeavor is the model's proficiency in accurately capturing the thermal wave behavior characteristic of hydrodynamic and ballistic conduction regimes, which cannot be replicated in models confined to a one-dissipative variable framework \cite{ZHAO2022122396}. Moreover, the model assures not only long-term stability but also exhibits an enhanced accuracy over a broad spectrum of Knudsen numbers in comparison to the typical Guyer-Krumhansl (G-K) model. The model's performance, particularly with discontinuous initial conditions—despite being calibrated solely on smooth initial conditions—further underscores its versatility.

In addition, we should point out that there are many ``schools" \cite{2007Why} for modeling heat conduction, such as Classical Irreversible Thermodynamics (CIT) \cite{De2013}, Extended Irreversible Thermodynamics (EIT) \cite{jou1996extended}, General Equation for Non-equilibrium Reversible-Irreversible Coupling (GENERIC) \cite{pavelka2018multiscale}. The CDF allows more freedoms which can also be determined by the deep neural networks, and thus, we choose CDF as the modeling method in this work, which is clarified in our previous paper \cite{ZHAO2022122396}.

\begin{figure}[htb]
  \centering
  \includegraphics[width=0.55\linewidth]{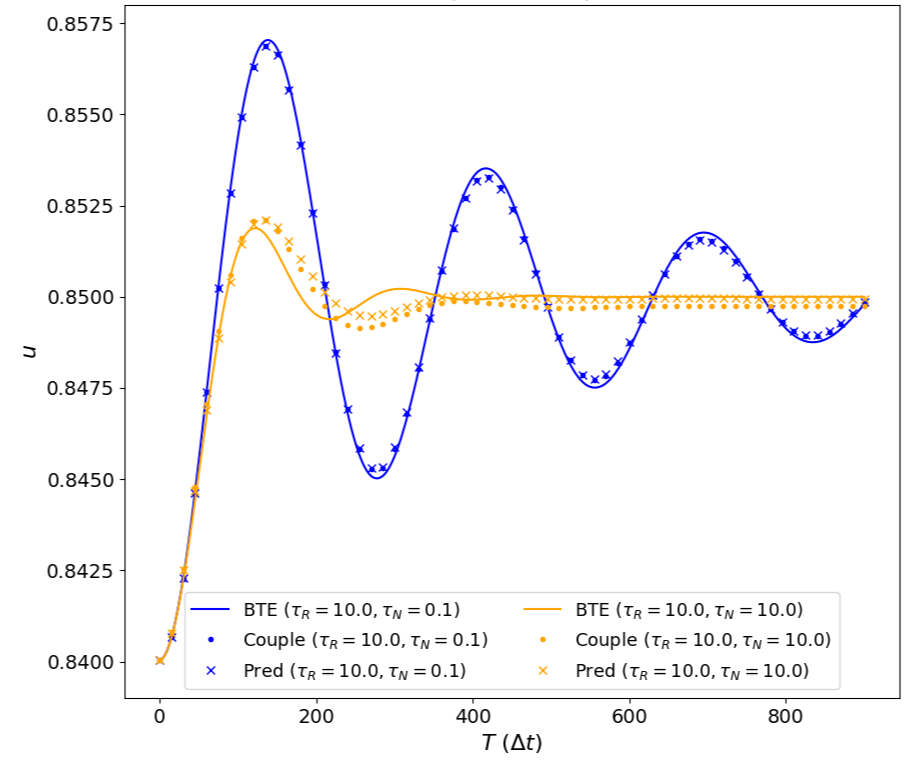}
  \caption{The numerical results $u(-\pi+30\Delta t)$ of \eqref{D5.1} and \eqref{2.5} about $\tau_R=10,~\tau_N=0.1$ and $\tau_R=10,~\tau_N=10$.}\label{coupled model}
\end{figure}
\begin{figure}[htb]
\centering
\subfigure[]{
\begin{minipage}[t]{0.45\linewidth}
\centering
\includegraphics[width=2.8in]{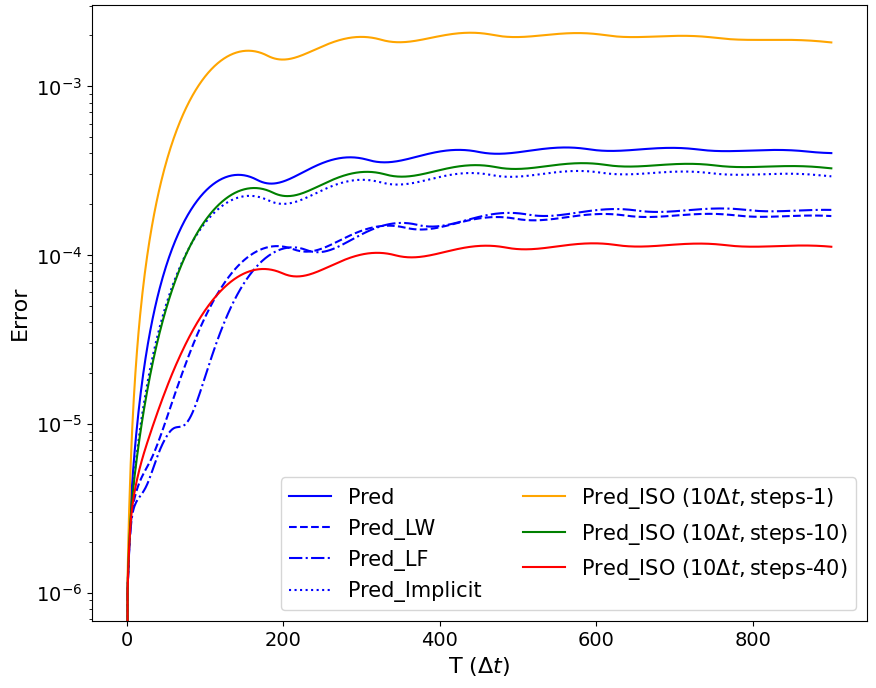}
\end{minipage}%
}%
\subfigure[]{
\begin{minipage}[t]{0.45\linewidth}
\centering
\includegraphics[width=2.8in]{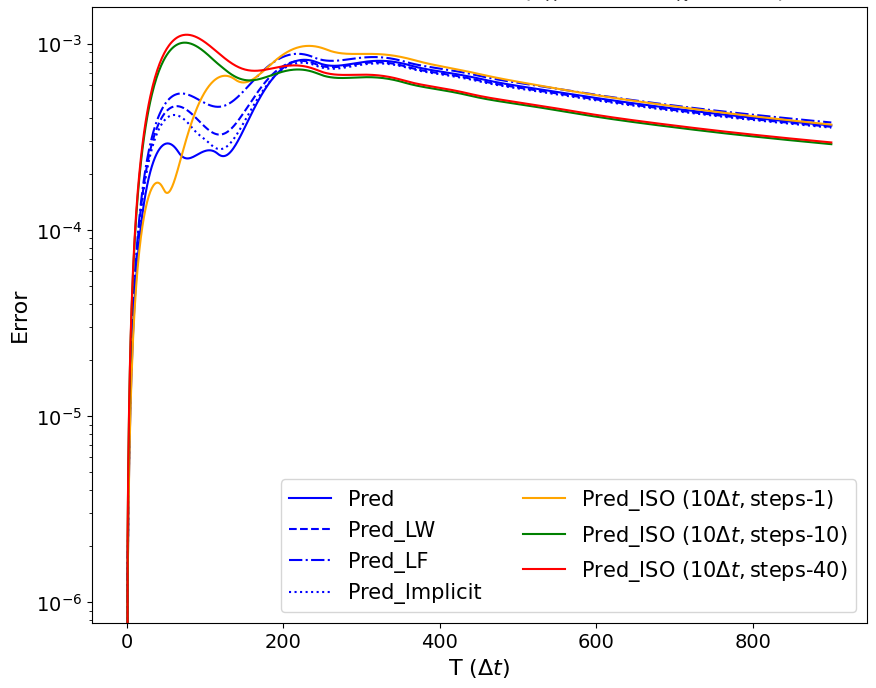}
\end{minipage}%
}%
\centering
\caption{The mean$-L^2$ relative errors of $u$ with the original version, Lax-Wendroff (LW) scheme, Lax-Friedrichts (LF) scheme, implicit scheme and the ISO with $10\Delta t$ about (a) $\tau_R=10,~\tau_N=0.1$ and (b) $\tau_R=10,~\tau_N=10$.}\label{OtherDiscrete}
\end{figure}

\subsection{Coupled Model}
It is noticed that \eqref{2.3} is not unique, and we also investigate the coupled model as follows:
\begin{equation}\label{D5.1}
\begin{aligned}
    \partial_t u + \partial_x q = & 0, \\
    \partial_t {w} + \partial_x \theta^{-1}+\gamma\partial_x s_{Q} = &{M'}_{11}q + {M'}_{12}s_{Q},\\
    \partial_t {Q} + \gamma\partial_x {q} = &{M'}_{21}q + {M'}_{22}s_{Q}.
\end{aligned}
\end{equation}
Here $$M'=M'(u,q,Q)=\left(
                     \begin{array}{cc}
                       {M'}_{11} & {M'}_{12} \\
                       {M'}_{21} & {M'}_{22} \\
                     \end{array}
                   \right)
$$
is a positive matrix, and thus \eqref{D5.1} satisfies the entropy law. Fig. \ref{coupled model} shows the numerical results of \eqref{D5.1} with the comparison of \eqref{2.5}. It is concluded that the specific model \eqref{2.5} illustrates the ballistic and hydrodynamic non-Fourier heat conduction very well enough.

\subsection{Other Discrete Versions}

The proposed model is trained by the discrete version \eqref{2.6}, which is based on a discrete time version, and thus some other discrete versions can be studied. The results about original version, Lax-Wendroff scheme, Lax-Friedrichts scheme, implicit scheme and the ISO with less data are plotted in Fig. \ref{OtherDiscrete}. This shows that higher order discrete versions seem not to make great improvements to the results.
In addition, the training method of the continuous time version is also used for \eqref{2.5}, such as physics-informed neural network (PINN) \cite{raissi2019physics}. However, we don't achieve an acceptable result for the time being, and perhaps it is left for future investigations.

\section{Conclusions $\&$ Remarks}
\label{sec:conclusion}

In this paper, we propose a mechanism-data fusion method (MDFM) for modeling heat conduction with dual-dissipative variables.
This method inherits advantages of mathematical rigor and adaptable machine learning, and further this model can be solved by conventional numerical methods directly.
Specifically, we use the conservation-dissipation formalism (CDF) to derive an interpretable system of partial differential equations (PDEs) for heat conduction, which obeys the first and second laws of thermodynamics.
The PDEs are macroscopic heat conduction equations for all phonon transport regimes, which are not limited by small Knudsen numbers.
Next, we train the unknown functions in this PDE system with deep neural networks {\color{black}(DNNs)}; this involves a ``warm-up" technique which prepares the connection of several time series.
Moreover, we propose a method, the inner-step operation (ISO), to diminish the extra errors caused by the selected discrete version.
Lots of numerical tests are conducted to show that the proposed model can well predict the heat conduction in diffusive, hydrodynamic and ballistic regimes, the last two of which can not be captured by the model with one-dissipative variable \cite{ZHAO2022122396}.
The model displays long-term stability, and demonstrates higher accuracy under a wider range of Knudsen numbers than the Guyer-Krumhansl (G-K) model.

{\color{black}In this paper, we select DNNs for our modeling due to their substantial computational capabilities and flexibility. Notably, one of the most advantageous features of DNNs is their capacity for transfer learning \cite{Tan2018DeepTransfer}, which enables them to adapt seamlessly to new data. This adaptability makes them especially suitable for dynamic environments where data continuously evolve, thereby enhancing the model's performance and ensuring its sustained relevance. Future updates to our model will leverage this capacity to further refine its accuracy and expand its applicability across varying datasets.}
In addition, it is remarkable that due to simplicity and the strictness of the training data, we only consider quasi-one dimensional case in this current work. Actually, the derivation \eqref{2.2} is generic for multi-dimensional cases, but it does not contain the boundary condition which is necessary for the multi-dimensional cases. Investigation into modeling for heat conduction in multi-dimensions, including presenting a compatible boundary condition, is our on-going and future work, and in turn, the result of the current work indicates that the MDFM is reasonable and powerful, which opens up a feasible and new way for modeling method.
----------------------------------------------------------------------------------------
\section*{Data Availability}
The data, including training data and testing data, have been deposited in the public netdisk (\url{https://disk.pku.edu.cn/link/AAF733D3A09571462A9D55C29127541EC6}) without any restrictions.
\section*{Code Availability}
The codes used in this study have been deposited in the public Github ({\url{https://github.com/LehengChen/HeatModelMDFM}}) without any restrictions.
\section*{Acknowledgements}
We thank Yuhao Hu and Bin Dong for useful discussions. 
This work is supported by the National Natural Science Foundation of China with grant No. 12301520 and No.12147122.
\section*{References}

\bibliographystyle{elsarticle-num-names_clear}
\bibliography{ref}

\end{document}